 \documentclass[10pt,preprint]{aastex}

\shorttitle{Star Formation in AGN}
\shortauthors{Davies et al.}

\begin{document}

\title{A Close Look at Star Formation around Active Galactic
    Nuclei\altaffilmark{1}}

\author{R. I. Davies, F. Mueller S\'anchez, R. Genzel, L.J. Tacconi,
  E.K.S. Hicks, S. Friedrich,} 
\affil{Max Planck Institut f\"ur extraterrestrische Physik, Postfach 1312,
  85741, Garching, Germany}

\and

\author{A. Sternberg}
\affil{School of Physics and Astronomy, Tel Aviv University, Tel
  Aviv 69978, Israel}



\altaffiltext{1}{Based on 
    observations at the European Southern Observatory VLT (60.A-9235, 
    070.B-0649, 070.B-0664, 074.B-9012, 076.B-0098).}

\begin{abstract}
We analyse star formation in the nuclei of 9
Seyfert galaxies at spatial resolutions down to 0.085\arcsec,
corresponding to length scales of order 10\,pc in most objects.
Our data were taken mostly with the near infrared adaptive optics
integral field spectrograph SINFONI.
The stellar light profiles typically have size scales of
a few tens of parsecs.
In two cases there is unambiguous kinematic evidence for stellar disks
on these scales.
In the nuclear regions there appear to have been recent -- but no longer
active -- starbursts in the last 10-300\,Myr.
The stellar luminosity is less than a few percent of the
AGN in the central 10\,pc, 
whereas on kiloparsec scales the luminosities are comparable.
The surface stellar luminosity density follows a similar trend in
all the objects, increasing steadily at smaller radii up to 
$\sim10^{13}$\,L$_\odot$\,kpc$^{-2}$ in the central few parsecs, where
the mass surface density exceeds 10$^4$\,M$_\odot$\,pc$^{-2}$.
The intense starbursts were probably Eddington limited and hence
inevitably short-lived, implying that the starbursts occur in
multiple short bursts.
The data hint at a delay of 50--100\,Myr between the onset of star
formation and subsequent fuelling of the black hole.
We discuss whether this may be a consequence of the role that stellar
ejecta could play in fuelling the black hole.
While a significant mass is ejected by OB winds and supernovae, their high
velocity means that very little of it can be accreted. 
On the other hand winds from AGB stars ultimately dominate the total mass
loss, and they can also be accreted very efficiently because of their slow
speeds.
\end{abstract}


\section{Introduction}
\label{sec:intro}

During recent years there has been increasing evidence for a
connection between active galactic nuclei (AGN) and star formation in
the vicinity of the central black holes.
This subject forms the central topic of this paper, and is discussed in
Sections~\ref{sec:prop} and ~\ref{sec:starAGN}.

A large number of studies have addressed the issue of star formation
around AGN.
Those which have probed closest to the nucleus, typically on scales of a
few hundred parsecs, have tended to focus on Seyferts -- notably
Seyfert\,2 galaxies -- since these are the closest examples
\citep{sar07,asa07,gon05,cid04,gon01,gu01,jog01,sto01,iva00}.
The overall conclusion of these studies is that in 30--50\% cases
the AGN is associated with young (i.e. age less than a few 100\,Myr)
star formation.
While this certainly implies a link, it does not necessarily imply any
{\em causal} link between the two phenomena.
Instead, it could more simply be a natural consequence of the fact
that both AGN and starburst require gas to fuel them.
And that in some galaxies this gas has fallen towards the
nucleus, either due to an interaction or secular evolution such as bar
driven inflow.

One aspect which must be borne in mind when interpreting such results,
and which has been pointed out by \cite{kna04}, is the discrepancy in
the scales involved.
AGN and starburst phemonena occur on different temporal and spatial scales; and
observations are sensitive to scales that are different again.
For example, star formation has typically been studied on scales of
several kiloparsecs down to a few hundred parsecs.
In contrast, accretion of gas onto an AGN will
occur on scales much less than 1\,pc.
Similarly, the shortest star formation timescales that most observations are
sensitive to are of order 100\,Myr to 1\,Gyr.
On the other hand, in this paper we show that the active phase of star
formation close around a black hole is typically rather less than
100\,Myr.
Correspondingly short accretion timescales for black holes are reflected 
in the ages of jets which, for a sample of radio galaxies measured by
\cite{mac07}, span a range from a few to 100\,Myr.
In Seyfert galaxies the timescales are even shorter, as typified by
NGC\,1068 for which \cite{cap99} estimate the age of the jets to be only
$\sim0.1$\,Myr.
That the putative causal connection between AGN and starbursts might occur on
relatively small spatial scales and short timescales can help us to
understand why no correlation has been found between AGN and
(circum-)nuclear starbursts in general.
It is simply that the circumnuclear activity on scales greater than a
few hundred parsecs is, in most cases, too far from the AGN to influence
it, or be strongly influenced by it \citep[cf][]{hec97}.

In this paper we redress this imbalance.
While the optical spectroscopy pursued by many authors allows a
detailed fitting of templates and models to the stellar features, 
we also make use of established star formation diagnostics and
interpret them using starburst population synthesis models.
Observing at near infrared wavelengths has brought two important
advantages.
The optical depth is 10 times less than
at optical wavelengths, and thus our data are less prone to the effects
of extinction which can be significant in AGN.
And we have employed adaptive optics to reach spatial resolutions of
0.1--0.2\arcsec, bringing us closer to the nucleus.
Applying these techniques,
we have already analysed the properties of the nuclear star formation
in a few objects \citep{dav04a,dav04b,dav06,mul06}.
Here we bring those data together with new data on 5 additional objects.
Our sample enables us to probe star
formation in AGN from radii of 1\,kpc down to less than 10\,pc.
Our aim is to ascertain whether there is evidence for star
formation on the smallest scales we can reach; and if so, to constrain
its star formation history. 
Ultimately, we look at whether there are indications that the nuclear
starburst and AGN are mutually influencing each other.

In \S\ref{sec:obs} we describe the sample selection, observations,
data reduction, PSF estimation, and extraction of the emission and
absorption line morphologies and kinematics.
In \S\ref{sec:diag} we discuss the observational diagnostics
and modelling tools.
Brief analyses of the relevant facets of our new data for the
individual objects are provided in 
Appendix~\ref{sec:obj}, where we also summarise results of our previously 
published data, re-assessing them where necessary to ensure that all
objects are analysed in a consistent manner.
The primary aims of our paper are addressed in \S\ref{sec:prop} and
\S\ref{sec:starAGN}.
In \S\ref{sec:prop} we discuss global results concerning the existence
and recent history of nuclear star formation for our whole sample.
In \S\ref{sec:starAGN} we discuss the implications of nuclear
starbursts on the starburst-AGN connection.
Finally, we present our conclusions in \S\ref{sec:conc}.

\section{Sample, Observations, Data Processing}
\label{sec:obs}

\subsection{Sample Selection}

The AGN discussed in this paper form a rather heterogeneous group. 
They include type~1 and type~2 Seyferts, ULIRGs, and even a QSO, and
do not constitute a complete sample.
In order to maximise the size of the sample, we have combined objects
on which we have already published adaptive optics near infrared
spectra with new observations of additional targets.

\begin{figure}
\epsscale{0.5}
\plotone{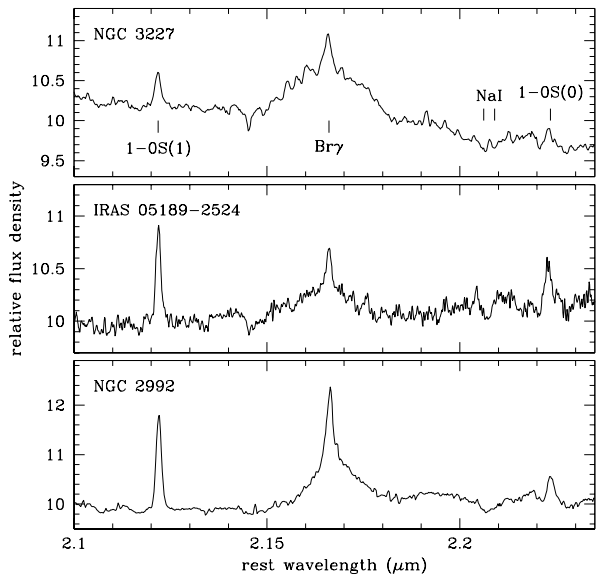} 
\caption{K-band spectra showing broad Br$\gamma$ emission in 3 
  AGN which are not usually classified as Seyfert~1.
Top: NGC\,3227 (0.25\arcsec\ aperture);
Middle: IRAS\,05189$-$2524 (1\arcsec\ aperture);
Bottom: NGC\,2992 (0.5\arcsec\ aperture).
The most prominent emission and absorption features are marked.}
\label{fig:bbrg}
\epsscale{1.0}
\end{figure}

Source selection was driven largely by technical considerations for
the adaptive optics (AO) system, namely having a nucleus bright and
compact enough to allow a good AO correction.
This is actually a strength since it means that 7 of the 
9 AGN are in fact broad line objects -- as given either by the standard
type~1 classification or because there is clear broad 
(FWHM $>1000$\,km\,s$^{-1}$) Br$\gamma$ emission in our spectra.
Fig.~\ref{fig:bbrg} shows broad Br$\gamma$ in K-band spectra of 3 AGN
that are not usually classified as broad line galaxies.
This is in contrast to most other samples of AGN for which star
formation has been studied in detail, and avoids any bias that might
arise from selecting only type~2 Syeferts.
That there may be a bias arises from the increasing evidence that the
obscuration in perhaps half of type~2 AGN lies at kpc scales rather
than in the nucleus, which may be caused by spatially extended star
formation in the galaxy disk \citep{bra07,mar06,rig06}.
Such AGN do not fit easily into the standard unification scheme (and
perhaps should not really be considered type~2 objects).
Because broad lines can be seen in the infrared, we know that we are
seeing down to the nuclear region and hence our results are not
subject to any effects that this might otherwise introduce.

It is exactly broad line AGN for which little is known about the nuclear star
formation, because the glare of the AGN swamps any surrounding stellar light
in the central arcsec.
As a result, most studies addressing star formation close to AGN have
focussed on type~2 Seyferts.
Adaptive optics makes it possible to confine much of the AGN's 
light into a very compact region, and to resolve the stellar continuum
around it.
The use of adaptive optics does give rise to one
difficulty when attempting to quantify the results in a uniform way,
due to the different resolutions achieved -- which is a
combination of both the distance to each object (i.e. target
selection) and the AO performance.
As a result, the standard deviation around the logarithmic mean
resolution of our sample (excluding NGC\,2992, see
Section~\ref{sec:obj}) of 22\,pc is a factor of 3.
However, this has enabled us to study the centers of AGN across nearly
3 orders of magnitude in spatial scale, from 1\,kpc in the more distant
objects to only a few parsecs in the nearby objects with the best AO
correction.

\subsection{Observations \& Reduction}

A summary of our observations is given in Table~\ref{tab:obs}.
A description of observations and processing of the new data is given
below.

Data for IRAS\,05189-2524 and NGC\,1068 were taken in December 2002 at the VLT
with NACO, an 
adaptive optics near infrared camera and long slit spectrograph
\citep{len03,rou03}.
Since IRAS\,05189-2524 is nearly face on, there is no strongly preferred axis
and the slit was oriented north-south;
for NGC\,1068 two orientations were used, north-south and east-west.
In all cases the slit width was 0.086\arcsec, yielding a nominal resolution of
$R\sim1500$ with the wide-field camera (pixel scale 0.054\arcsec) and medium
resolution grism. 
The galaxy was nodded back and forth along the slit by 10\arcsec\ to allow sky
subtraction.
For IRAS\,05189-2524, 12 integrations of 300\,sec were made;
for NGC\,1068 12 integrations of 300\,sec were made at one position angle, and
14 frames of 200\,sec at the other.
All data were reduced and combined, using standard longslit techniques in
IRAF, to make the final H-band spectrum.

Data for NGC\,7469, NGC\,2992, NGC\,1097, NGC\,1068, and NGC\,3783 were taken
during 2004--2005 at the VLT with SINFONI, an adaptive optics near infrared
integral field spectrograph \citep{eis03,bon04}.
Data were taken with various gratings covering the H and K bands either
separately ($R\sim4000$) or together ($R\sim1500$).
The pixel scales were 0.125\arcsec$\times$0.25\arcsec\ or
0.05\arcsec$\times$0.1\arcsec, depending on the trade-offs between field of
view, spatial resolution, and signal-to-noise ratio.
Individual exposure times are in the range 50--300\,sec depending on
the object brightness.
Object frames were interspersed with sky frames, usually using the
sequence O-S-O-O-S-O, to facilitate background subtraction.
The data were processed using the dedicated {\it spred} software package
\citep{abu06},
which provides similar processing to that for longslit data but with the added
ability to reconstruct the datacube.
The data processing steps are as follows.
The object frames are pre-processed by subtracting sky frames,
flatfielding, and correcting bad pixels (which are identified from
dark frames and the flatfield).
The wavemap is generated, and edges and curvature of the slitlets are
traced, all from the arclamp frame.
The arclamp frame is then reconstructed into a cube, which is checked
to ensure that the calibration is good.
The pre-processed object frames are then also reconstructed into
cubes, spatially shifted to align them using the bright nucleus as a
reference, and combined.
In some cases the final cube was spatially smoothed using a $3\times3$
median filter. 
Estimation of the spatial resolution (see below) was always performed
after this stage.

In some cases, the strong near-infrared OH lines did not subtract
well.
With longer exposure times this is to be expected since the timescale
for variation of the OH is only 1--2\,mins.
If visual inspection of the reconstructed cubes showed signs of over-
or under-subtraction of the OH lines, these cubes were reprocessed
using the method described in \cite{dav07a}.

Standard star frames are similarly reconstructed into cubes.
Telluric correction and flux calibration were performed using B stars (K-band)
or G2V stars (H-band).
In addition, flux calibration was cross-checked in 3\arcsec\ apertures
using 2MASS data, and in smaller 1--3\arcsec\ apertures using
broad-band imaging from NACO or HST NICMOS.
Agreement between cubes with different pixel scales, and also with the
external data, was consistent to typically 20\%.

\subsection{PSF Estimation}

There are a multitude of ways to derive the point spread
function (PSF) from adaptive optics data, five of which are
described in \cite{dav07b}.
With AGN, it is usually possible to estimate the PSF from the science
data itself, removing any uncertainty about spatial and temporal
variations of the PSF due to atmospheric effects.
Typically one or both of the following methods are employed on the
new data presented here.
If a broad emission line is detected, this will always yield a measure
of the PSF since the BLR of Seyfert galaxies has a diameter that can
be measured in light days.
Alternatively, the non-stellar continuum will provide a sufficiently
good approximation in all but the nearest AGN since at near infrared
wavelengths it is expected to originate from a region no more than
1--2\,pc across.

In every case we have fit an analytical function to the PSF. 
Since the Strehl ratio achieved is relatively low,
even a Gaussian is a good representation.
We have used a Moffat function, which achieves a better fit because it 
also matches the rather broad wings that are a characteristic of
partial adaptive optics correction.
The PSF measured for NGC\,3227, which is shown in Fig.~1 of
\cite{dav06}, can be considered typical.
If one applies the concept of `core plus halo' to this PSF, then the
Gaussian fit would represent just the core while the Moffat fit the
entire `core plus halo'.
Integrating both of these functions indicates that about 75\% of the
flux is within the `core', and it is thus this component which
dominates the PSF.
In this paper, a more exact representation of the PSF is not needed since
we have not performed a detailed kinematic analysis, and
we have simply used the Moffat to derive a FWHM for the spatial resolution.
The resolutions achieved are listed in Table~\ref{tab:obs}.

\subsection{Emission/Absorption Line Characterisation}

The 2D distribution of emission and absorption features has been found by
fitting a function to the continuum-subtracted spectral profile at
each spatial position in the datacube.
The function was a convolution of a Gaussian with a spectrally
unresolved template profile -- in the case of emission lines it was an
OH sky emission line, and for stellar absorption features we made use
of template stars observed in the same configuration (pixel scale and
grism).
A minimisation was performed in which the parameters of the Gaussian
were adjusted until the convolved profile best matched the data.
During the minimisation, pixels in the data that consistently
deviated strongly from the data were rejected.
The uncertainties were boot-strapped using Monte Carlo techniques,
assuming that the noise is uncorrelated and the intrinsic profile is
well represented by a Gaussian.
The method involves adding a Gaussian with the derived properties
to a spectral segment that exhibits the same noise statistics
as the data, and refitting the result to yield a new set of Gaussian
parameters.  After repeating this 100 times, the standard deviation of
the center and dispersion were used as the uncertainites for the
velocity and line width.

The kinematics were further processed using kinemetry \citep{kra06}.
This is a parameterisation (i.e. a mathematical rather than a physical
model) of the 2D field.
As such, beam smearing is not a relevant issue to kinemetry, which
yields an analytical expression for the observed data.
Of course, when the coefficients of this expression are interpreted or
used to constrain a physical model, then beam smearing should be considered.
Mathematically, the kinemetry procedure fits the data with a linear
sum of sines
and cosines with various angular scalings around ellipses at each radius. 
We have used it for 3 purposes: to determine the best position angle
and axis ratio for the velocity field, to remove high order noise
from the raw kinematic extraction, and to recover the velocity and
dispersion radial profiles.
In all of the cases considered here, the kinematic centre of the
velocity field was assumed to be coincident with the peak of the
non-stellar continuum. 
In addition, the uniformity of the velocity
field permitted us to make the simplifying assumption of a single
position angle and axis ratio --
i.e. there is no evidence for warps or twisted velocity contours.
We then derived the position angle and inclination of the
disk by minimising the {\em A1} and {\em B3} parameters respectively
(see \citealt{kra06} for a description of these).
The rotation curves were recovered by correcting the measured velocity
profile for inclination.
We have assumed throughout the paper that the dispersion is isotropic,
and hence no inclination correction was applied to the
dispersion that was measured.

The innermost parts of the kinematics derived as above are of course
still affected by beam smearing.
In general, the central dispersion cannot necessarily be taken at face
value since it may either be artificially increased by any component of
rotation included within the beam size, or decreased if neighbouring
regions within the beam have a lower dispersion.
In the galaxies we have studied, there are two aspects which mitigate
this uncertainty: 
the rotation speed in the central region is much less than the
dispersion and so will not significantly alter it;
and when estimating the central value we consider the trend of the
dispersion from large radii, where the effect of the beam is small, to
the center.
For the basic analyses performed here, we have therefore adopted the
central dispersion at face value.
More detailed physical models for the nuclear disks, which properly
account for the effects of beam smearing, will be presented in future
publications.
Lastly, we emphasize that the impact of the finite beam size on the
derived rotation curve does not affect our measurement of the
dynamical mass.
The reason is that, for all the dynamical mass estimates we
make, the mass is estimated at a radius much large than the FWHM of
the PSF -- as can be seen in the relevant figures.

\section{Quantifying the Star Formation}
\label{sec:diag}

In this section we describe the tools of the trade used to analyse
the data, and which lead
us to the global results presented in Section~\ref{sec:prop}.
Specific details and analyses for individual objects can be found in
Appendix~\ref{sec:obj}.
We use the same methods and tools for all the objects
to ensure that all the data are analysed in a consistent manner.

Perhaps the most important issue is how to isolate the stellar
continuum, which is itself a powerful diagnostic.
In addition, we use three standard and independent diagnostics to
quantify the star formation history and intensity in the nuclei of
these AGN.
These are the Br$\gamma$ equivalent width, supernova rate, and
mass-to-light ratio.
Much of the discussion concerns how we take into account the
contribution of the AGN when quantifying these parameters.
We also consider what impact an incorrect compensation could
have on interpretation of the diagnostics.

We model these observational diagnostics using the stellar population
and spectral synthesis code STARS
\citep[e.g.][]{ste98,ste03,for03,dav03,dav05}.
This code calculates the distribution of
stars in the Hertzsprung-Russell diagram as a function of age for an
assumed star formation history.
We usually assume an exponentially decaying star formation rate, which
has an associated timescale $\tau_{\rm SF}$.
Spectral properties of the cluster are then computed given the
stellar population present at any time.
We note that the model output of STARS is quantitatively similar to
that from version 5.1 of Starburst99, which unlike earlier versions
does include AGB tracks \citep{lei99,vaz05}.
As discussed in detail below, particular predictions of STARS include,
for ages greater than 10\,Myr: 
equivalent widths of $W_{\rm CO2-0}\sim12\AA$ and 
$W_{\rm CO6-3}\sim4.5\AA$, and H-K color of 0.15\,mag.
The equivalent quantities predicted by Starburst99 v5.1 are 
$W_{\rm CO2-0}\sim11\AA$ and $W_{\rm CO6-3}\sim5\AA$, 
and H-K color of 0.2\,mag.

\subsection{Isolating the stellar continuum}

\begin{figure}
\epsscale{0.32}
\plotone{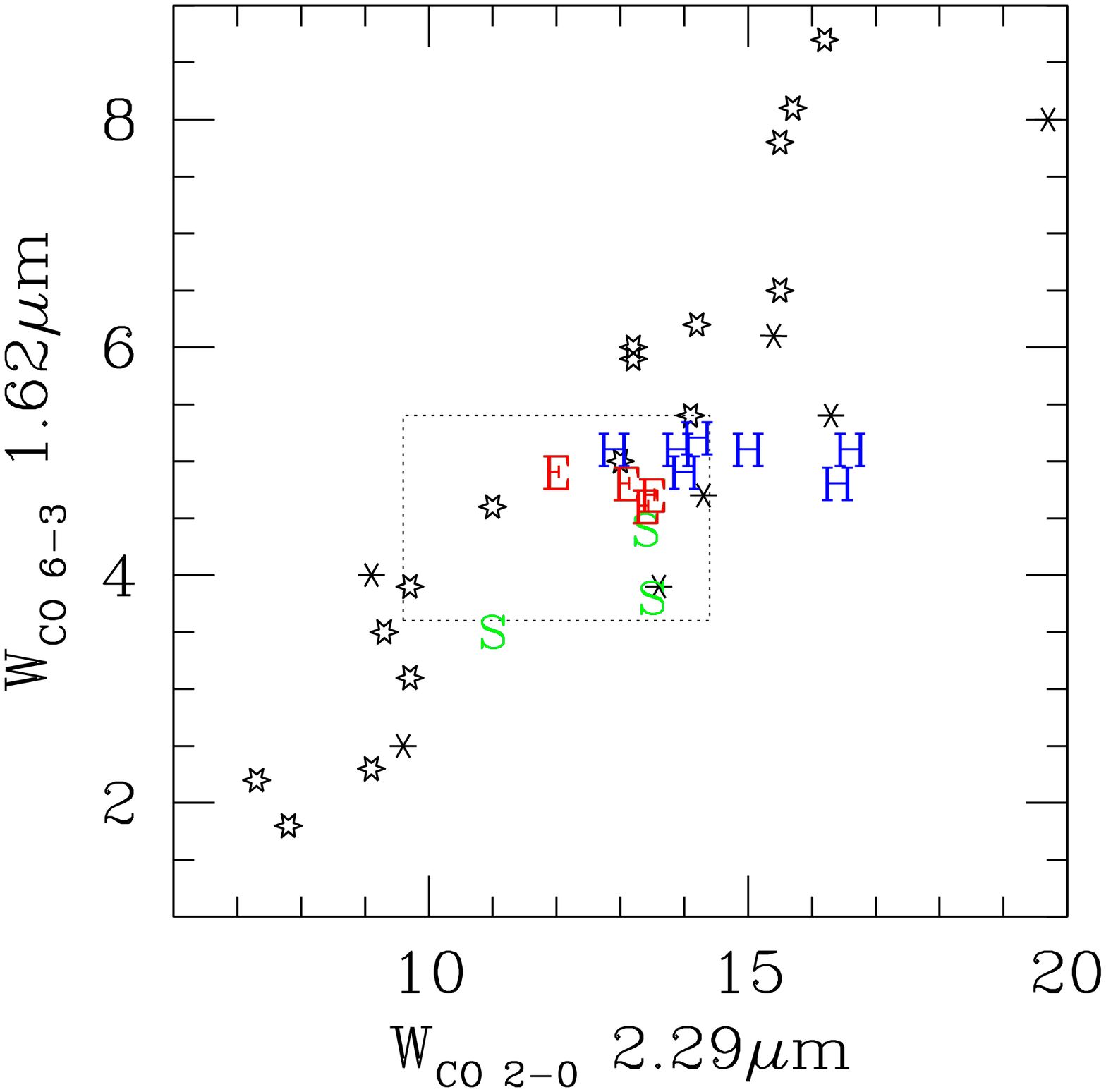} 
\plotone{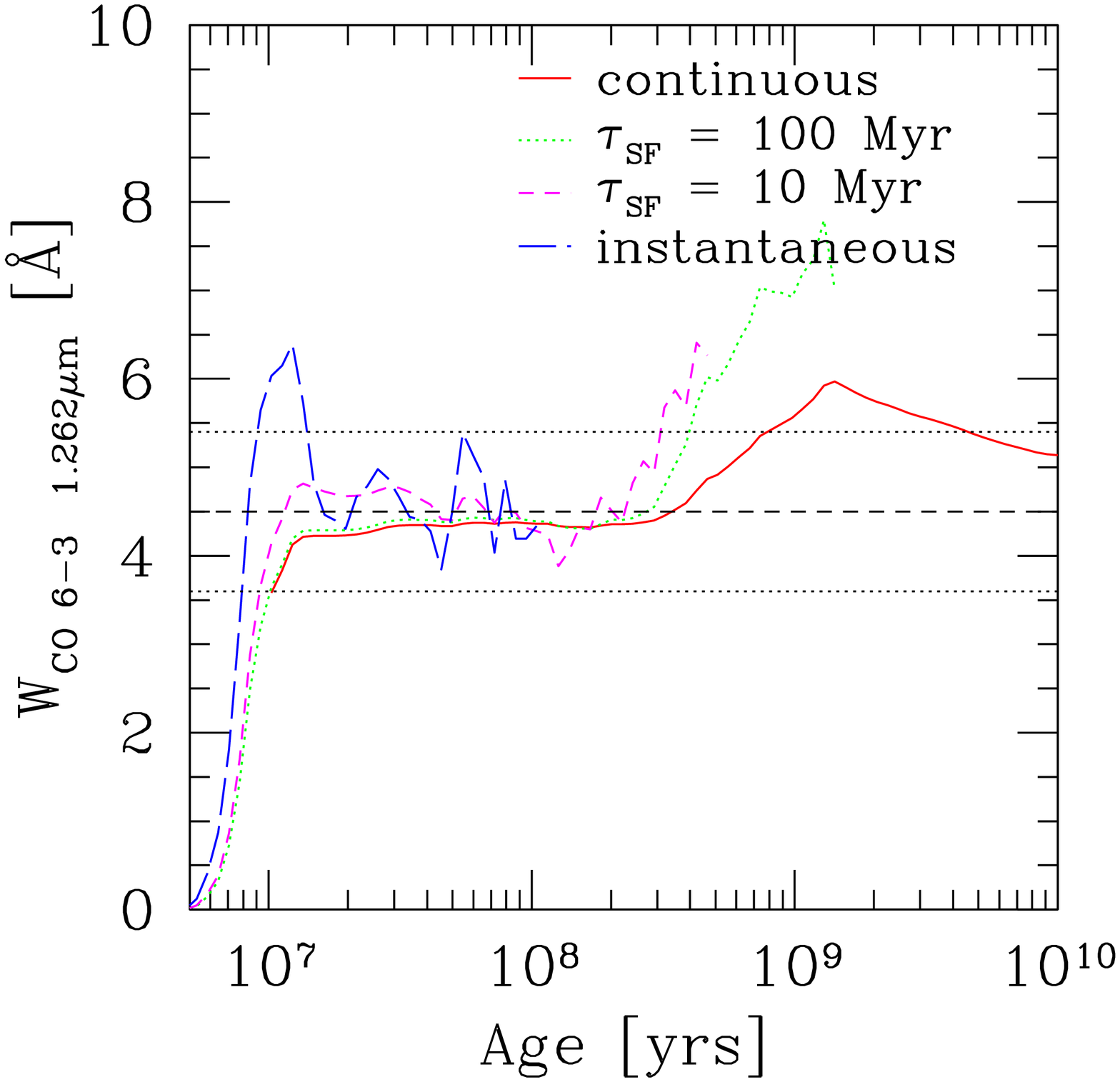} 
\plotone{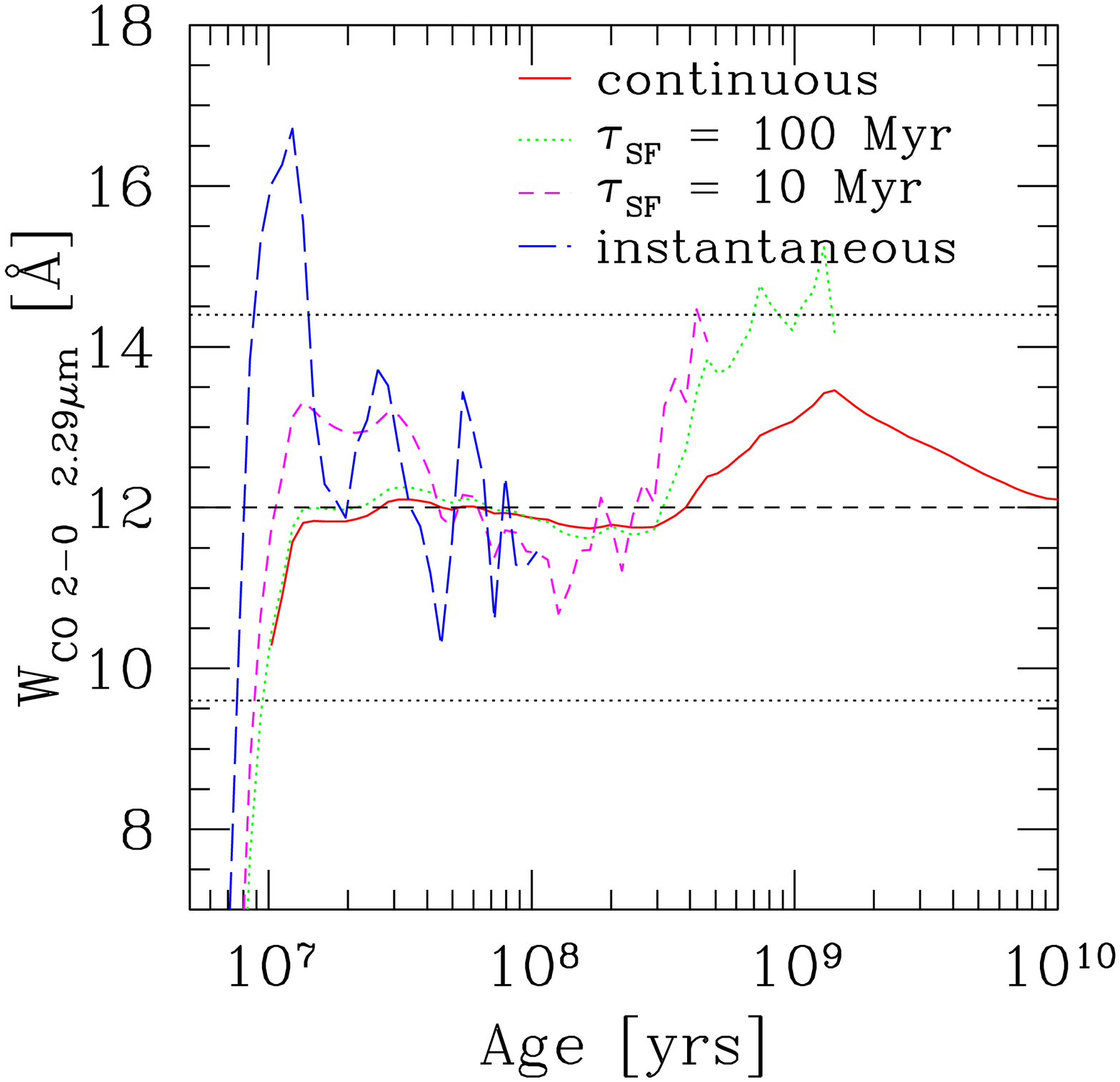} 
\caption{Left: equivalent width of the CO\,2-0 and CO\,6-3 features for
  various stars and galaxies.
The late-type supergiant stars (skeletal star shapes) and giant stars
  (open star shapes) are taken from \cite{ori93}.
The galaxies (denoted `E' for elliptical, `S' for spiral, and `H' for star
  forming H{\sc ii} galaxy) are from \cite{oli95}.
The dashed box encloses the region for which there is no more than
  20\% deviation from each of the values $W_{\rm CO2-0}=12$\AA\ 
 and $W_{\rm CO6-3}=4.5$\AA.
Centre and right: calculated $W_{\rm CO6-3}$ and $W_{\rm CO2-0}$
  respectively from STARS for several different star formation
  histories. Each line is truncated when the cluster luminosity falls
  below 1/15 of its maximum. In each case, the dashed lines show
  typical values adopted, and the dotted lines denote a range of $\pm$20\%.}
\label{fig:equiv}
\epsscale{1.0}
\end{figure}

For small observational apertures a significant fraction of the 
K-band (and even H-band) 
continuum can be associated with non-stellar AGN continuum.
The AGN contribution can be estimated from a simple measurement of the
equivalent width of a stellar absorption feature.
We use CO\,2-0 2.29\micron\ in the K-band or CO\,6-3 1.62\micron\ in
the H-band.
Although the equivalent widths $W_{\rm CO2-0}$ and $W_{\rm CO6-3}$
vary considerably for individual stars, the integrated values for stellar
clusters span only a rather limited range.
This was shown by \cite{oli95} who measured these values for
elliptical, spiral, and star-forming (H{\sc ii}) galaxies.
We have plotted their measurements of these two absorption features in
the left-hand panel of 
Fig.~\ref{fig:equiv}, together with the equivalent widths of giant and
supergiant stars from \cite{ori93}.

In STARS we use empirically
determined equivalent widths from library spectra \citep{for00} to
compute the time-dependent equivalent width for an entire
cluster of stars.
Results for various star formation histories are shown in the centre
and right panels of Fig.~\ref{fig:equiv}, for $W_{\rm CO6-3}$ 
and $W_{\rm CO2-0}$ respectively.
Typical values are $W_{\rm CO6-3}=4.5$\AA\ and $W_{\rm CO2-0}=12$\AA.
The dashed box in the left panel shows that the locus of 20\% deviation from
each of these computed values is consistent with observations. 
That the H{\sc ii} galaxies have slightly higher $W_{\rm CO2-0}$ can
be understood because these are selected to have bright emission lines
and hence are strongly biassed towards young stellar ages -- often
corresponding to the maximum depth  
of the stellar features that occurs at 10\,Myr due to the late-type
supergiant population.
It may be this bias for galaxies selected as `starbursts', and the
similarity of the CO depth for starbursts 
of all other ages, that led \cite{iva00} to conclude that there is no
evidence for strong starbursts in Seyfert\,2 galaxies.
Similarly, an estimate of the dilution can be
found from the Na{\sc i}\,2.206\micron\ line.
Fig.~7 of \cite{dav05} shows that for nearly all star formation
histories the value $W_{\rm Na\,I}$ remains in the range 2--3\AA.

Our conclusion here is that within a reasonable uncertainty of
$\pm$20\% (see Fig.~\ref{fig:equiv}), one can
assume that the intrinsic equivalent width of the absorption -- most
notably CO -- features of any stellar population that contains late-type stars
is independent of the star formation history and age.
For a stellar continuum diluted by additional non-stellar emission,
the fraction of stellar light is 
\[
f_{\rm stellar} = W_{\rm obs} / W_{\rm int}
\]
where $W_{\rm obs}$ and $W_{\rm int}$ are the observed and intrinsic
equivalent widths of the CO features discussed above.
Thus, we are able to correct the observed continuum magnitude for the
contribution associated with the AGN.

\subsection{Stellar colour and luminosity}

\begin{figure}
\epsscale{0.32}
\plotone{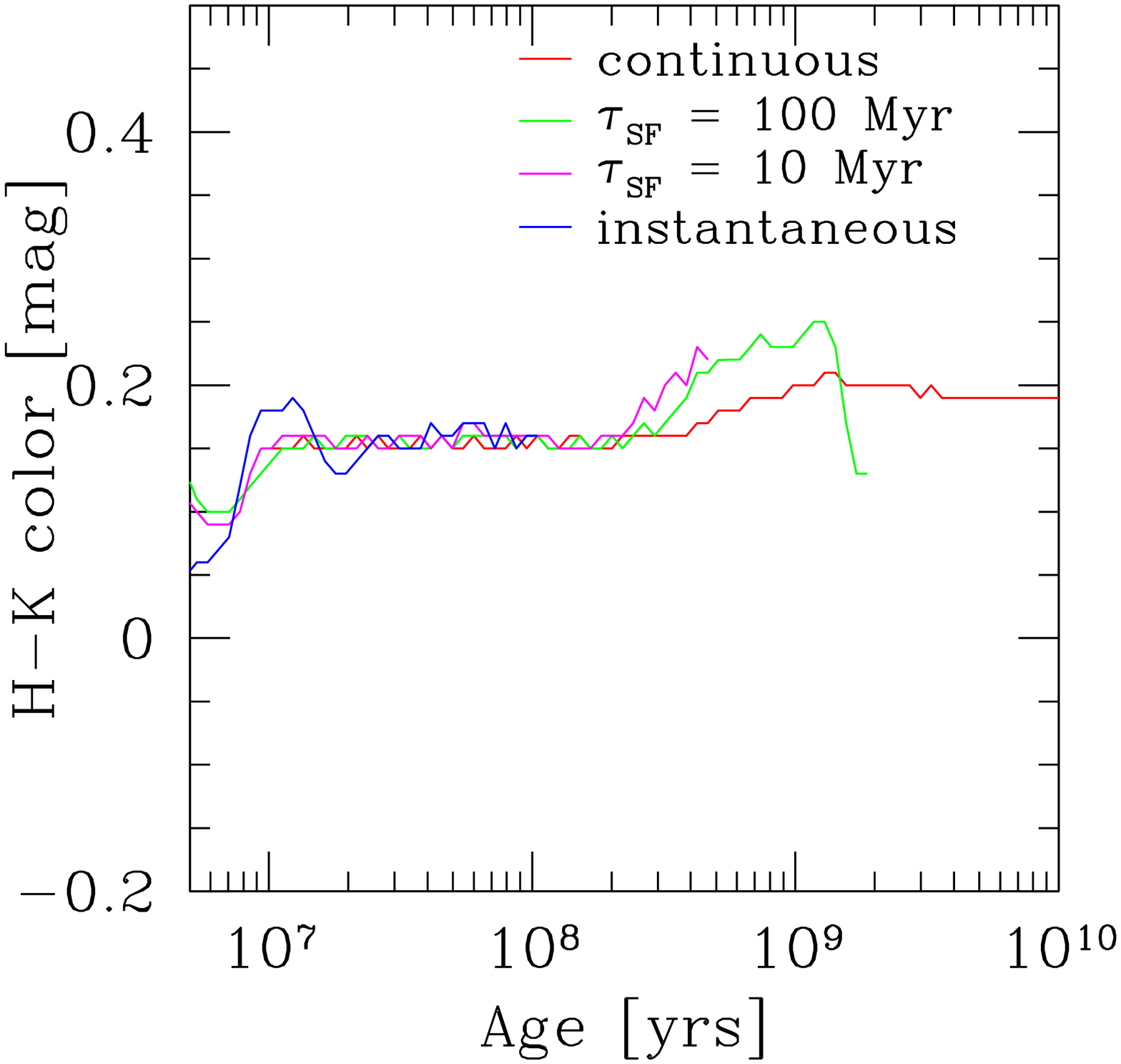} 
\plotone{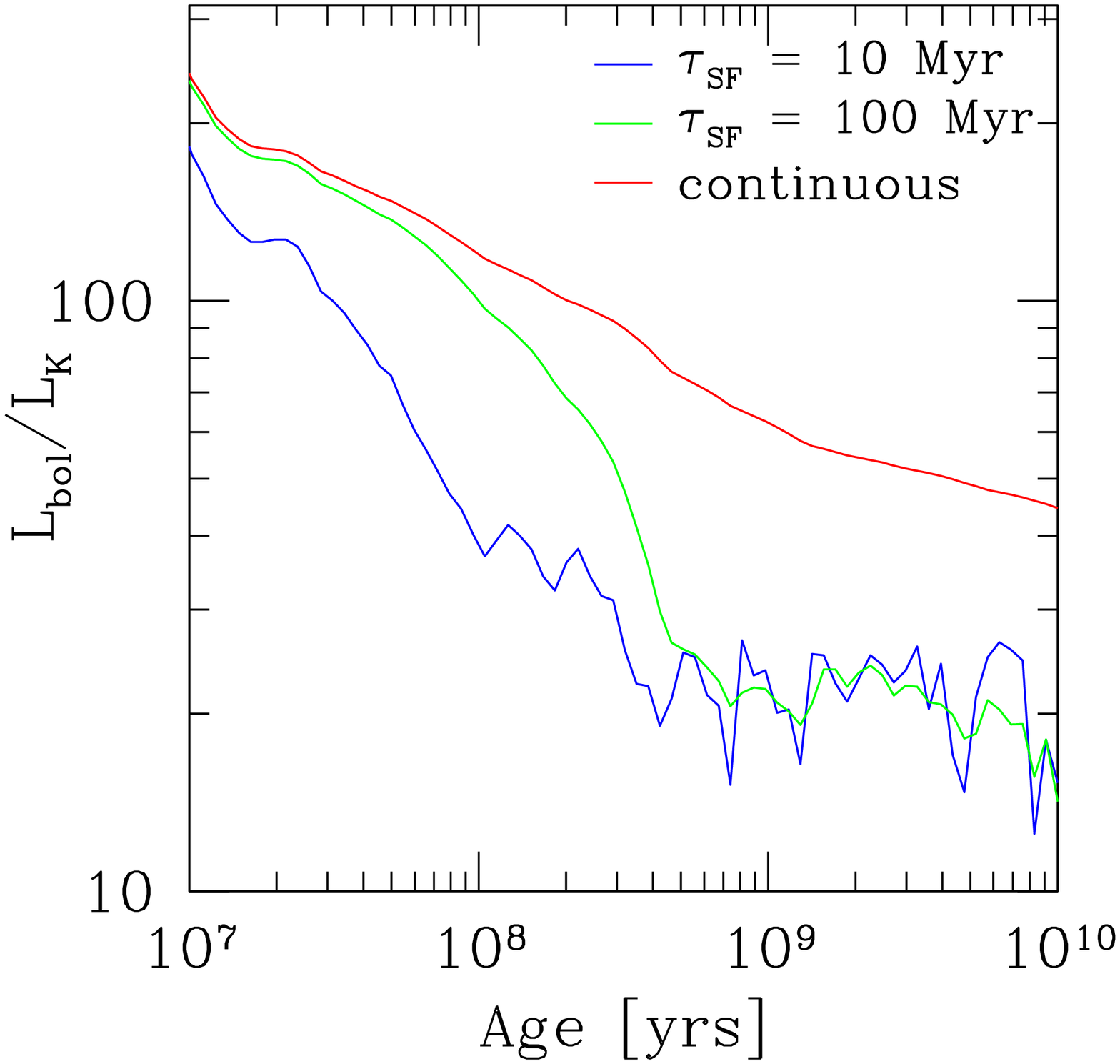} 
\caption{Left: H-K colour of star clusters with different star
  formation timescales and ages, as calculated by STARS.
Right: Ratio of bolometric to K-band luminosity. Although the range of
  20--200 initially appears large, the uncertainty on an intermediate
  value of 60 is only 0.3 \,dex. This is small compared to the range
  of interest in the paper, which is several orders of
  magnitude.}
\label{fig:col_lum}
\epsscale{1.0}
\end{figure}

Our data cover both the H and K-bands -- hence the reason for using
both $W_{\rm CO6-3}$ and $W_{\rm CO2-0}$.
In order to homogenize the dataset, we need to convert H-band
stellar magnitudes to K-band.
The STARS computation in Fig.~\ref{fig:col_lum} shows that this
conversion is also independent of the star formation history, being
close to $H-K=0.15$\,mag (no extinction) for all timescales and ages.
This result is supported empirically by photometry of elliptical and
spiral galaxies performed by \cite{gla84}.
For ellipticals $H-K\sim0.2$--0.25, and for spirals $H-K\sim0.2$--0.3.
Some of the difference between the data and models could be due to
extinction since $H-K = (H-K)_0 + (A_H - A_K)$; 
and for $A_V=1$,  $A_H - A_K = 0.08$.
However, at the level of precision required here, the 5--10\%
difference between model and data can be considered negligible.

To convert from absolute magnitude to luminosity we use the relation 
\[
M_K = -0.33 - 2.5\log{L_K}
\]
where $L_K$ is the total luminosity in the
1.9--2.5\micron\ band in units of bolometric solar luminosity
($1\,L_\odot = 3.8\times10^{26}$\,W), and as such different from
the other frequently used monochromatic definition with 
units of the solar K-band luminosity density 
($2.15\times10^{25}$\,W\,$\mu$m$^{-1}$).
We then use STARS to estimate the bolometric
stellar luminosity $L_{\rm bol}$.
The relation between $L_{\rm bol}$ and $L_K$ is shown in the right
panel of Fig.~\ref{fig:col_lum}. 
The dimensionless ratio $L_{\rm bol}/L_K$ depends on the age and the
exponential decay timescale of the star formation.
However, the range spanned is only 20--200 for ages greater than
10\,Myr.
Thus even if the star formation history cannot be
constrained, a conversion ratio of $L_{\rm bol}/L_K \sim 60$ will have
an associated uncertainty of only 0.3\,dex.
In general we are able to apply constraints on the star formation age,
and so our errors will be accordingly smaller.

\subsection{Specific Star Formation Diagnostics}

Graphs showing how the diagnostics vary with age and star formation
timescale are shown in Fig.~\ref{fig:stars}.

\subsubsection{Br$\gamma$ equivalent width}

Once the stellar continuum luminosity is known, an upper limit to the
equivalent width of Br$\gamma$ asssociated with star formation can be
found from the narrow Br$\gamma$ line flux.
In some cases it is possible to estimate what fraction of the narrow
Br$\gamma$ might be associated with the AGN.
This can be done both morphologically, for example if
the line emission is extended along the galaxy's minor axis; and/or
kinematically, for example if the line shows regions that are broader,
perhaps with FWHM a few hundred km\,s$^{-1}$, suggestive of outflow.
Even if acounting for the AGN contribution is not possible, one may be
able to set interesting upper limits or even rule out continuous star
formation scenarios, and put a constraint on the time since
the star formation was active.
This can be seen in the lefthand panel of Fig.~\ref{fig:stars}, which
shows for example that for ages less than $10^9$\,yrs, continuous star
formation scenarios will always have $W_{\rm Br\gamma}>12$\,\AA.

\subsubsection{Supernova rate}

We estimate the type {\sc ii} (core collapse) supernova rate
$\nu_{SN}$ from the radio continuum using the relation \citep{con92}:
\[
L_N (W\,Hz^{-1}) \,\, = \,\,
           1.3\times10^{23} \,\,\,
           \nu^{-\alpha} (GHz) \,\,\, 
           \nu_{SN} (yr^{-1})
\]
where $L_N$ is the non-thermal radio continuum luminosity, $\nu$ is
frequency of the observation and $\alpha\sim0.8$ the spectral index of
the non-thermal continuum.
This relation was derived for Galactic supernova remnants; but a
similar one, differing only in having a coefficient of
$1.1\times10^{23}$, was derived by \cite{hua94} for M\,82.
For the 5\,GHz non-thermal radio continuum luminosity of Arp\,220
\citep[176\,mJy,][]{ana00} it would lead to a supernova rate of
2.9\,yr$^{-1}$, comfortably within the 1.75--3.5\,yr$^{-1}$ range
estimated by \cite{smi98} based on the detection of individual
luminous radio supernovae.
This, therefore, seems a reasonable relation to apply to starbursts.

We have to be careful, however, to take into account any
contribution from the AGN to the radio continuum.
Our premise for the nuclei of Seyfert galaxies is that if the nuclear
radio continuum is spatially resolved 
(i.e. it has a low brightness temperature) and does not have the
morphology of a jet, it is likely to
originate in extended star formation.
At the spatial scales of a few parsec or more that we can resolve,
emission from the AGN will be very compact.
As a result, we can use the peak surface brightness to estimate the
maximum (unresolved) contribution from an AGN.
Wherever possible, we use radio continuum observations at a comparable
resolution to our data to derive the extended emission; 
and observations at higher resolution to estimate the AGN
contribution.
Details of the data used in each case are given in the relevant
sub-sections for each object in Appendix~\ref{sec:obj}.
In addition, we exclude any emission obviously associated with jets,
for example as in NGC\,1068.

To use $\nu_{SN}$ as a diagnostic, we normalise it with respect to the
stellar K-band luminosity.
This gives the ratio $10^{10}\nu_{SN}/L_K$,
for which STARS output is drawn in Fig.~\ref{fig:stars}.

\subsubsection{Mass-to-light ratio}

Models indicate that the ratio $M/L_K$ of the stellar mass to K-band
luminosity should be an excellent diagnostic since, for ages greater
than 10\,Myr, it increases monotonically with age as shown in
Fig.~\ref{fig:stars}.

However, in practice estimating the stellar mass is not entirely
straightforward.
In many cases it is only practicable to derive the dynamical mass.
It may be possible to estimate and hence correct for the molecular gas
mass based on millimetre CO maps, but these are scarce at sufficiently
high spatial resolution and are associated with their own CO-to-H$_2$
conversion uncertainties.
We also note that it is often not possible to separate the `old' and
`young' stellar populations.
The best one can do is estimate the overall mass-to-light
ratio, and argue that this is an upper limit to the true ratio for
the young population.
While there inevitably remains uncertainty on the true
ratio, the limit is often sufficient to apply useful constraints on
the age of the `young' population.

Our estimates of the dynamical mass are based wherever possible on the
stellar kinematics, since the gas kinematics can be perturbed by
warps, shocks, and outflows.
We begin by estimating the simple Keplerian mass assuming that the
stars are supported by ordered rotation at velocity 
$V_{\rm rot} = V_{\rm obs}/\sin{i}$ in a thin plane.
However, the stellar kinematics in all the galaxies exhibit a
significant velocity dispersion indicating that a considerable mass is
supported by random rather than ordered motions.
Thus the simple Keplerian mass is very much an underestimate, and
any estimate of the actual mass is associated with large 
uncertainties -- see for example \cite{ben92}, who derive masses of
spheroidal systems. 
As stated in Section~\ref{sec:obs}, we assume that the random motions
are isotropic. 
Our relation for estimating the mass enclosed within a radius $R$ is then
\[
M = (V_{\rm rot}^2 + 3\sigma^2) R / G.
\]
where $\sigma$ is the observed 1-dimensional velocity dispersion.

We note that when taking rotation into account in estimating the
masses of spheroids with various density profiles, \cite{ben92} also
use a factor 3 between the $V$ and $\sigma$ terms in their 
Appendix B.
Despite the complexities involved, within the unavoidable uncertainties
(a factor 2--3), their relation gives the same mass as that above.
Although this uncertainty appears to be quite large, it does not
impact the results and conclusions in this paper since we are
concerned primarily with order-of-magnitude estimates when considering
mass surface densities.

\section{Properties of Nuclear Star Formation}
\label{sec:prop}

In the following section we bring to together the individual results
(detailed in Appendix~\ref{sec:obj}) to form a global picture.
It is possible to do this because all the data have been analysed in a
consistent manner, using the tools described in Section~\ref{sec:diag}
to compare in each object the same diagnostics to the same set
of stellar evolutionary synthesis models.

We note that the discussion that follows is based on results for 8 of
the AGN we have observed.
As explained in Appendix~\ref{sec:obj} we exclude NGC\,2992 because we
are not able to put reliable constraints on the properties of the
nuclear star formation.
Despite this, there are indications that at higher spatial resolution one
should expect to find a distinct nuclear stellar population as has
been seen in other AGN.

\paragraph{Size scale}

Tracing the stellar features rather than the broad-band continuum, we have in
all cases resolved a stellar population in the nucleus close around the AGN.
While this should not be unexpected if the stellar distribution follows a
smooth $r^{1/4}$ or exponential profile, we have in several cases been
able to show that on scales of $<50$\,pc there is in fact an excess
above what one would expect from these profiles.
This suggests that in general we are probing an inner star forming
component.

Fig.~\ref{fig:sizes} shows normalised azimuthally averaged stellar luminosity
profiles for the AGN.
These have not been corrected for a possible old underlying
population, nor has any deconvolution with the PSF been performed.
Nevertheless, it is still clear that the stellar intensity increases very
steeply towards the nucleus.
In 6 of the 8 galaxies shown, the half-width at half-maximum is less than
50\,pc.
The remaining 2 galaxies are the most distant in the sample, and the
spatial resolution achieved does not permit a size measurement on
these scales.
We may conclude that the physical radial size scale of the nuclear star
forming regions in Seyfert galaxies does not typically exceed 50\,pc.

\paragraph{Stellar Age}

For 8 of the AGN studied here, we have been able to use classical star
formation diagnostics based on line and continuum fluxes as well as
kinematics to constrain the ages of the inner star forming regions.
The resulting ages should be considered `characteristic', since in
many cases there may simultaneously be two or more stellar populations
that are not co-eval.
For example, if a bulge population exists on these small spatial
scales, it was not usually possible to account for the contamination
it would introduce.
While this would have little effect on $W_{\rm Br\gamma}$, it could
impact on $M/L_{\rm K}$ more strongly, increasing the inferred age.
The ages we find lie in the range 10--300\,Myr, compelling evidence
that it is common for there to be relatively young star clusters
close around AGN.

Intriguingly, we also find rather low values of $W_{\rm Br\gamma}$:
typically $W_{\rm Br\gamma}\lesssim10$\,\AA\ (see
Table~\ref{tab:derprop}).
This indicates directly that there is currently little or no on-going
star formation.
Coupled with the relatively young ages, we conclude that the star
formation episodes are short-lived.
One may speculate then that the star formation is episodic, recurring
in short bursts.
The scale of the bursts and time interval between them would certainly
have an impact on the fraction of Seyfert nuclei in which
observational programmes are able to find evidence for recent star
formation.

\paragraph{Nuclear Stellar Disks}

The first evidence for nuclear stellar disks came from seeing limited
optical spectroscopy, for which a slight reduction in $\sigma_*$ was seen for
some spiral galaxies \citep{ems01,marq03,sha03}.
And there is now a growing number of spiral galaxies -- more than 30
-- in which the phenomenon has been observed, suggesting that they
might occur in 30\% or more of disk galaxies \citep{ems06c}.
The $\sigma_*$-drop has been interpreted by \cite{ems01} as arising from a
young stellar population that is born from a dynamically cold gas
component, and which makes a significant contribution to the total
luminosity.
This appears to be borne out by N-body and SPH simulations of
isolated galaxies \citep{woz03}, which suggest that although the entire
central system will slowly heat up with time, the $\sigma_*$-drop can
last for at least several hundred Myr.
Indeed, preliminary analysis of optical integral field data for
NGC\,3623 suggest that the stellar population responsible for the
$\sigma_*$-drop cannot be younger than 1\,Gyr \citep{ems06b}.

Our results provide strong support for the nuclear disk interpretation.
In previous work \citep{dav06,mul06}, we had argued that in both
Circinus and NGC\,3227 the inner distributions were disk-like, albeit
thickened.
We have now found much more direct evidence for this phenomenon in
NGC\,1097 and NGC\,1068.
In both of these galaxies, we have spatially resolved a
$\sigma_*$-drop and an excess stellar continuum over the same size
scales.
In NGC\,1097 this size was $\sim$0.5\arcsec, corresponding to about 40\,pc.
For NGC\,1068 these effects were measured out to $\sim$1\arcsec,
equivalent to 70\,pc.
These are not the scale lengths of the disks, but simply the maximum
radius to which we can detect them.
In both cases the mean mass surface densities are of order 
$\Sigma = $(1-3)$ \times10^4$\,M$_\odot$\,pc$^{-2}$.
For an infinitely large thin self gravitating stellar disk, one can use the
expression  $\sigma_z^2 = 2 \pi G \Sigma z_0$
to estimate the scale height.
Although this may not be entirely appropriate, we use it here to obtain a
rough approximation to the scale heights, which are 5--20\,pc.
Thus while the disks appear to be flattened, they should still be
considered thick since the radial extent is only a few times the scale
height.

The impact of nuclear starbursts on the central light
profile of galaxies was considered theoretically more than a decade
ago by \cite{mih94}.
They performed numerical simulations of galaxy mergers to study the
mass and luminosity profiles of the remnants, 
taking gas into account, and estimating the star formation rate
using a modified Schmidt law.
They found that there should be a starburst in the nucleus which would
give rise to an excess stellar continuum above the $r^{1/4}$ profile
of the older stars in the merged system.
Several years ago, compact nuclei were found to be present in a
significant fraction of spiral galaxies \citep{bal03} as well as Coma
cluster dwarf ellipticals \citep{gra03}.
More recently, 
nuclei with a median half-light radius of 4.2\,pc have been found in
the majority of early-type members of the Virgo Cluster \citep{cot06};
and traced out to $\sim$1\arcsec, equivalent to $\sim100$\,pc, in some
of the `wet' merger remnants in that cluster \citep{kor07}.
While the nuclear starbursts in these latter cases are caused by a merger
event, whereas those we are studying arise from secular evolution as gas
from the galaxy disk accretes in the nucleus, there appear to be many
parallels in the phenomenology of the resulting starbursts.

\paragraph{Star Formation Rate}

It is possible to estimate the bolometric luminosity $L_{\rm bol*}$ of
the stars from their K-band luminosity $L_{\rm K}$ even if one knows
nothing about the star formation history.
As discussed in Section~\ref{sec:diag}, this would result in an
uncertainty of about a factor 3.
The diagnostics in Table~\ref{tab:derprop} and discussions in
Appendix~\ref{sec:obj} enable us to apply some constraints to the
characteristic age of the star formation.
Because continuous star formation is ruled out by the low 
$W_{\rm Br\gamma}$, we have assumed 
exponential decay timescales of $\tau_{\rm SF}=10$--100\,Myr.
We have then used STARS to estimate the average star formation rates.
In order to allow a meaningful comparison between the objects, the
rates have been normalised to the same area of 1\,kpc$^2$.
These are the rates given in Table~\ref{tab:derprop}.
They are calculated simply as the mass of stars produced divided by 
the entire time since the star forming episode began.
Because $\tau_{\rm SF}$ is shorter than the age, the average
includes both active and non-active phases of the starburst.
Indeed, for $\tau_{\rm SF}=10$\,Myr one would expect the star formation
  rate during the active phases to be at least a factor of a few, and
  perhaps an order of magnitude, greater.
The table shows that on scales of a few hundred parsecs one might
expect a few $\times10$\,M$_\odot$\,yr$^{-1}$\,kpc$^{-2}$, while on
scales of a few tens of parsecs mean rates reaching
$\sim100$\,M$_\odot$\,yr$^{-1}$\,kpc$^{-2}$ should not be unexpected;
and correspondingly higher -- up to an order of magnitude, see
Fig.~\ref{fig:starstoy} -- during active phases.

An obvious question is why there should be such vigorous star
formation in these regions.
Star formation rates of 10--100\,M$_\odot$\,yr$^{-1}$\,kpc$^{-2}$ are
orders of magnitude above those in normal galaxies and comparable to
starburst galaxies.
The answer may lie in the Schmidt law and the mass surface densities we have
estimated in Table~\ref{tab:derprop}.
Fig.~\ref{fig:msurfden} shows these surface densities at the radii
over which they were estimated, revealing a trend towards higher
densities on smaller scales and values of a few times
$10^4$\,M$_\odot$\,pc$^{-2}$ in the central few tens of parsecs.
The global Schmidt law, as formulated by \cite{ken98}, states that the star
formation rate depends on the gas surface density as 
$\Sigma_{\rm SFR} \propto \Sigma_{gas}^{1.4}$.
If one assumes that 10--30\% of the mass in our AGN is gas, then this
relation would predict time-averaged star formation rates in the range
10--100\,M$_\odot$\,yr$^{-1}$\,kpc$^{-2}$, as have been observed.
That the high star formation rates may simply be a consequence of the
high mass surface densities is explored futher by Hicks et al. (in prep.).

\paragraph{Stellar Luminosity}

As a consequence of the high star formation rates, the stellar
luminosity per unit area close around the AGN is very high in these
objects.
Despite this, because the star formation is occurring only in very
small regions, the absolute luminosities are rather modest.
This can be seen in Fig.~\ref{fig:mag_agn} which shows the bolometric
luminosity of the stars as a fraction of the entire bolometric
luminosity of the galaxy.
We have calculated a range for the ratio $L_{\rm bol*}/L_{\rm K}$
appropriate for each galaxy based on the ages in
Table~\ref{tab:derprop} for different $\tau_{\rm SF}$.
Because we assume that all the K-band stellar continuum is associated
with the young stars, we have adopted the lower end of each range in
an attempt to minimise possible overestimation of $L_{\rm bol*}$.
The resulting values for the ratio used span 30--130, within a factor
of 2 of the `baseline' value of 60 given in Section~\ref{sec:diag}.
In the central few tens of parsecs, young stars contribute a few
percent of the total.
But integrated over size scales of a few hundred parsecs,
this fraction can increase to more than 20\%.
On these scales, the star formation is energetically significant
when compared to the AGN.
Such high fractions imply that on the larger scales the extinction to the
young stars must be relatively low.
On the other hand, on the smallest scales where in absolute terms the stellar
luminosity is small, there could in general be
considerable extinction even at near infrared wavelengths.
In this paper we have not tried to account for extinction since it is very
uncertain.
The primary effect of doing so would simply be to increase the stellar
luminosity above the values discussed here.

Fig.~\ref{fig:mag_bol} shows the stellar bolometric luminosity $L_{\rm bol*}$
integrated as a function of radius.
All the curves follow approximately the same
trend, with the luminosity per unit area increasing towards smaller
scales and approaching $10^{13}$\,L$_\odot$\,kpc$^{-2}$ in the central
few parsecs.
This appears to be a robust trend and will not change significantly
even with large uncertainties of a factor of a few.
It is remarkable that the luminosity density of
$10^{13}$\,L$_\odot$\,kpc$^{-2}$ is that estimated by \cite{tho05} for
ULIRGs, which they modelled as optically thick starburst disks.
The main difference between the ULIRG model and the starbursts close around
AGN is the spatial scales on which the starburst occurs.

Based on this model, they argued that ULIRGs are radiating at the
Eddington limit for a starburst, defined as when the radiation
pressure on the gas and dust begins to dominate over self-gravity.
The limiting luminosity-to-mass ratio was estimated to be
$\sim500$\,L$_\odot$/M$_\odot$ by \cite{sco03}.
He argued that in a star cluster, once the upper end of the main
sequence was populated, the radiation pressure would halt further
accretion on to the star cluster and hence terminate the star
formation.
Following  \cite{tho05}, we apply this definition to the entire disk
rather than a single star cluster.
For the $10^{13}$\,L$_\odot$\,kpc$^{-2}$, this
implies a mass surface density of $2\times10^4$\,M$_\odot$\,pc$^{-2}$.
Comparing these quantities to the AGN we have observed, we find that on
scales of a few tens of parsecs they are an order of magnitude below
the Eddington limit.
On the other hand, we have already seen that the low 
$W_{\rm Br\gamma}$ indicates that there is little on-going star
formation and hence that the starbursts are short-lived.
This is important because short-lived starbursts fade very quickly.
As shown in Fig.~\ref{fig:starstoy}, 
for a decay timescale of $\tau_{\rm SF}=10$\,Myr, $L_{\rm bol*}$
will have decreased from its peak value by more than an order of
magnitude at an age of 100\,Myr.
Thus it is plausible -- and probably likely -- that while the star
formation was active, the stellar luminosity was an order of magnitude
higher.
In this case the starbursts would have been at, or close to, their
Eddington limit at that time.

The luminosity-to-mass ratio of $500$\,L$_\odot$/M$_\odot$ associated
with the Eddington limit is in fact one that all young starbursts
would exceed if, beginning with nothing, gas was accreted at the same
rate that it was converted into stars.
That, however, is not a realistic situation.
A more likely scenario, shown in Fig.~\ref{fig:ratio_bol}, is that the
gas is already there in the disk.
In this case, a starburst with a star-forming timescale of 100\,Myr
could never exceed $100$\,L$_\odot$/M$_\odot$.
To reach $500$\,L$_\odot$/M$_\odot$, the gas would need to be
converted into stars on a timescale $\lesssim10$\,Myr.
This timescale is independent of how much gas there is.
Thus, for a starburst to reach its Eddington limit, it must be very
efficient, converting a significant fraction of its gas 
into stars on very short $\sim10$\,Myr timescales.
This result is consistent with the prediction of the Schmidt law,
which states that disks with a higher gas surface
density will form stars more efficiently.
The reason is that the star formation
efficiency is simply 
$SFE = \Sigma_{\rm SFR}/\Sigma_{\rm gas} 
     \propto \Sigma_{\rm gas}^{0.4}$.
Thus, from arguments based solely on the Schmidt law and mass surface
density, one reaches the same conclusion that the
gas supply would be used rather quickly and the lifetime of the
starburst would be relatively short.

Summarising the results above, a plausible scenario could be as follows.
The high gas density leads to a high star formation rate, producing a
starburst that reaches its Eddington limit for a short time.
Because the efficiency is high, the starburst can only be active for a
short time and then begins to fade.
Inevitably, one would expect that the starburst is then dormant
until the gas supply is replenished by inflow.
This picture appears to be borne out by the observations presented here.

\section{Starburst-AGN Connection}
\label{sec:starAGN}

In the previous sections we have presented and discussed evidence that in
general there appears to have been moderately recent star formation on
small spatial scales around all the AGN we have observed.
Fig.~\ref{fig:age} shows the first empirical indication of a deeper
relationship between the star formation and the AGN.
In this figure we show the luminosity of the AGN, both in absolute
units of solar luminosity and also in relative units of its Eddington
luminosity $L_{\rm Edd}$, against the age of the most recent known
nuclear star forming episode.
Since the AGN luminosity is not well known, we have made the
conservative assumption that it is
equal to half the bolometric luminosity of the galaxy -- as may be the
case for NGC\,1068 (\citealt{pie94}, but see also \citealt{bla97}).
To indicate the expected degree of uncertainty in this assertion we have
imposed errorbars of a factor 2 in either direction, equivalent to
stating that the AGN luminosity in these specific objects is likely to
be in the range 25--100\% of the total luminosity of the galaxy.
The Eddington luminosity is calculated directly from the black hole
mass, for which estimates exist for these galaxies from reverberation
mapping, the M$_{\rm BH}-\sigma*$ relation, maser kinematics, etc.
These are listed in Table~\ref{tab:basicdata}.
For the age of the star formation, we have plotted the time since the most
recent known episode of star formation began, as given in
Table~\ref{tab:derprop}.
For galaxies where a range of ages is given, we have adopted these to
indicate the uncertainty;
the mean of these, $\sim\pm$30\%, has been used to estimate the
uncertainty in the age for the rest of the galaxies.
We note that these errorbars reflect uncertainties in characterising the
age of the star formation from the available diagnostics and also in
the star formation timescale $\tau_{\rm SF}$.
However, there are still many implicit assumptions in this process,
and we therefore caution that the actual errors in our estimation of
the starburst ages may be larger than that shown.

Conceding this, we do not wish to over-interpret the figure.
Keeping the uncertainties in mind, Fig.~\ref{fig:age} shows
the remarkable result that AGN which are radiating
at lower efficiency $\lesssim0.1$\,L/L$_{\rm Edd}$ are associated with
younger $\lesssim50$--100\,Myr starbursts;
while those which are more efficient $\gtrsim0.1$\,L/L$_{\rm Edd}$
have older $\gtrsim50$--100\,Myr starbursts.
If one were to add to this figure the Galactic Centre -- which is known to
have an extremely low luminosity 
(L/L$_{\rm Edd} < 10^{-5}$; \citealt{oze96,bag03}) and to have experienced a 
starburst $6\pm2$\,Myr ago \citep{pau06} -- it would be consistent
with the categories above.
The inference is that either there is a delay
between the onset of starburst activity and the onset of AGN activity,
or star
formation is quenched once the black hole has become active.

In Section~\ref{sec:prop} we argued that the starbursts are to some
extent self-quenching:
that very high star formation efficiencies are not sustainable over
long periods.
In addition, an intense starburst will provide significant heat input to
the gas, which is perhaps partially responsible for the typically high
gas velocity dispersions in these regions (Hicks et al., in prep.).
This itself could help suppress further star formation.
Heating by the AGN could also contribute to this process, 
and has been proposed as the reason why the molecular torus is
geometrically thick \citep{pie92,kro07}.
It is also used to modulate star
formation (at least on global scales) in semi-analytic models of
galaxy evolution \citep{gra04,spr05}.
While this is certainly plausible, it does not explain either why the star
formation in some galaxies with a lower luminosity AGN has
already ceased, nor why none of the AGN associated with younger starbursts
are accreting efficiently.

Instead we argue for the former case above, that efficient
fuelling of a black hole is associated with a starburst that is at
least 50--100\,Myr old.
It may be because of such a delay between AGN and starburst
activity that recent star formation is often hard to detect close to
AGN: the starburst has
passed its most luminous (very young) age, and is in decline while the AGN
is in its most active phase (see Fig.~\ref{fig:starstoy}).
This does not necessarily imply that the {\em a priori} presence of a
starburst is required before an AGN can accrete gas -- although it
seems inevitable  that one will occur as gas accumulates in the
nucleus.
Nor does it imply that all starbursts will result in fuelling a
black hole; indeed it is clear that there are many starbursts not associated
with AGN.
As we argue below, the crucial aspect may be the stellar ejecta associated
with the starburst;
and in particular, not just the mass loss rate, but the speed with which the
mass is ejected.

\paragraph{Winds from OB stars}

In the Galactic Center, \cite{oze96} proposed that it is the recent
starburst there that is limiting the luminosity of the black hole.
In this scenario, mechanical winds from young stars -- both the
outflow and the angular momentum of the gas (which is a consequence of
the angular momentum of the stars themselves) --
hinder further inflow.
The authors argued that almost none of the gas flowing into the central parsec
reached the black hole because of outflowing winds from
IRS\,16 and He\,{\sc i} stars in that region.
Detailed modelling of the Galactic Center region as a 2-phase medium was
recently performed by \cite{cua06}.
They included both the fast young stellar winds with velocities of
700\,km\,s$^{-1}$ \citep{oze97} and the slower winds of
$\sim200$\,km\,s$^{-1}$ \citep{pau01}; and 
also took into account the orbital angular momentum of the stars
\citep{pau01,gen03}, which had a strong influence on reducing the
accretion rate.
They found that the
average accretion rate onto the black hole was only
$\sim3\times10^{-6}$\,M$_\odot$\,yr$^{-1}$, although an intermittent cold flow
superimposed considerable variability onto this.
In contrast, the hypothetical luminosity \cite{oze96} estimate that Sgr\,A$^*$
would have if it could accrete all the
inflowing gas, would be 
$5\times10^{43}$\,erg\,s$^{-1}$, typical of Seyfert galaxies.
In principle this process could be operating in other galaxy nuclei
where there has been a starburst which extends to less than 1\,pc from
the central black hole. 
However, it cannot explain the timescale of the delay we have
observed, which is an order of magnitude greater than the main sequence
lifetime of OB and Wolf-Rayet stars.

\paragraph{Winds from AGB stars}

Stars of a few (1--8\,M$_\odot$) solar masses will
evolve on to the asymptotic giant branch (AGB) at the end of their
main sequence lifetimes.
The timescale for stars at the upper end of this range to reach this phase
is $\sim50$\,Myr, comparable to the delay apparent in
Fig.~\ref{fig:mag_bol}.
Since AGB stars are known to have high mass-loss rates, of order
$10^{-7}$--$10^{-4}$\,M$_\odot$\,yr$^{-1}$ at velocities of
10--30\,km\,s$^{-1}$ \citep{win03}, they may be prime candidates for
explaining the delay between starburst and AGN activity.
To quantify this, we consider how much of the mass in the wind could
be accreted by the central supermassive black hole.

The Bondi parameterisation of the accretion rate onto a point particle
for a uniform spherically symmetric geometry is given by \citep{bon52}
\[
\dot{M} \ = \ \frac{2 \pi \, G^2 \, M^2 \, \rho}{(V^2 + c^2_s)^{3/2}}
\]
where $M$ is the mass of the point particle moving through a gas
cloud, $V$ is the
velocity of the particle with respect to the cloud, $\rho$ is the
density of the cloud far from the point particle, 
and $c_s$ is the sound speed.
This approximation is still used to quantify accretion on to
supermassive black holes in models of galaxy evolution \citep{spr05},
even though it may be significantly inaccurate for realistic
(e.g. turbulent) media \citep{kru06}.
Here, it is sufficient to provide an
indication of the role that stellar winds may play in accretion onto a
central black hole.
The density of the stellar wind at a distance $R$ from the parent star
is given by
\[
\rho_{\rm wind} \ = \ \frac{\dot{M}_{\rm wind}}{4 \, R^2 \, V_{\rm wind}}
\]
In our case, $R$ is the distance from the star to the black hole.
One would therefore expect that the accretion rate on to the black
hole could be written as (see also \citealt{mel92})
\[
\dot{M}_{\rm BH} \ \sim \ 
  \frac{G^2 \, M^2_{\rm BH} \, \dot{M}_{\rm wind}}
       {(V^2_{\rm wind}+c^2_s)^{3/2} \, V_{\rm wind} \, R^2}
\]
This equation shows that $\dot{M}_{\rm BH} \propto V_{\rm wind}^{-4}$.
We have implicitly assumed that $V_{\rm wind}$ is greater than the
orbital velocity $V_{\rm orb}$ of the star from which it originates.
This is not the case for AGB winds, and so one reaches the limiting
case of $\dot{M}_{\rm BH} \propto V_{\rm orb}^{-4}$, where 
for the galaxies we have observed $V_{\rm orb} \sim 50$--100\,km\,s$^{-1}$.
This is still at least an order of magnitude less than the winds from
OB and Wolf-Rayet stars.
Thus, even though the mass loss rates from individual OB
and Wolf-Rayet stars are similar to those of AGB stars, the AGB winds
will fuel a black hole much more efficiently.
However, for slow stellar winds that originate close to a
$10^7$\,M$_\odot$ black hole, the equation breaks down because the
conditions of uniformity and spherically symmetry are strongly violated.
Indeed, the apparent accretion rate exceeds the outflow rate --
implying that essentially the entire wind can be accreted.
For AGB wind velocities of 10--30\,km\,s$^{-1}$, the maximum radius at
which the entire wind 
from a star in Keplerian orbit around a $10^7$\,M$_\odot$ black hole
will not exceed the escape velocity from that orbit 
(i.e. $V_{\rm wind} + V_{\rm orb} < V_{\rm esc}$) 
is around 10--70\,pc.
We adopt the middle of this range, 40\,pc, as the
characteristic radius within which it
is likely that a significant fraction, and perhaps most, of the AGB
winds are accreted onto the black hole.
Fig.~\ref{fig:mag_bol} indicates that the stellar luminosity within
this radius is $\sim2\times10^9$\,L$_\odot$.
It is this luminosity that has been used to scale the STARS model (for
$\tau_{\rm SF}=10$\,Myr and an age of 100\,Myr) in
Fig.~\ref{fig:starstoy}, and so one can also simply read off the mass loss
from the figure.
The mass loss rate for such winds peaks at about 0.1\,M$_\odot$\,yr$^{-1}$ and
then tails off proportionally to the K-band luminosity, leading to a
cumulative mass lost of $2\times10^7$\,M$_\odot$ after 1\,Gyr (although most
of the loss occurs actually occurs within half of this timespan).
This mass loss rate is sufficient to power a Seyfert nucleus for a short time.
A typical Seyfert with M$_{\rm BH} \sim 10^7$\,M$_\odot$ requires
0.02\,M$_\odot$\,yr$^{-1}$ to radiate at the Eddington limit.
Even for the short bursts we have modelled, Fig.~\ref{fig:starstoy} shows that
this can be supplied by AGB winds for starburst ages in the range
50--200\,Myrs.

We note that taking an AGB star luminosity of $10^4$\,L$_\odot$ (which is at
the high 
end of the likely average, \citealt{nik97}) we then find that there are
$\sim2\times10^5$ AGB stars close enough to the black hole to
contribute to accretion.
In order to provide at least 0.02\,M$_\odot$\,yr$^{-1}$, the typical mass loss
rate per star must exceed $10^{-7}$\,M$_\odot$\,yr$^{-1}$, which is the lower
limit of the range measured for Galactic AGB stars given above.
Thus the mass losses and rates estimated here appear to be plausible.

The low speed of these winds means they will not create much turbulence. 
We quantify this by considering their total mechanical energy
$\frac{1}{2}mv^2$ integrated over the same timespan, which is
$\sim10^{45}$\,J.
These two quantities -- gas mass ejected and mechanical energy -- are
compared to those for supernovae below.

\paragraph{Supernovae}

Type {\sc ii} supernovae are the stellar outflows most able to create
turbulence in the interstellar medium, since they typically eject
masses of $\sim5$\,M$_\odot$ at 
velocities of $\sim5000$\,km\,s$^{-1}$ \citep{che77}.
Each supernova therefore represents a considerable injection of
mechanical momentum and energy into the local environment.
A large number of compact supernova remnants are known, for example in
M\,82 and Arp\,220, and are believed to have expanded into dense
regions with $n_{\rm H} \sim 10^3$--$10^4$\,cm$^{-3}$ \citep{che01}.
These authors argue that such remnants become radiative when they
reach sizes of $\sim1$\,pc, at which point
the predicted expansion velocity will have slowed to
$\sim500$\,km\,s$^{-1}$.
By this time, the shock front will have driven across
$\sim1000$\,M$_\odot$ of gas.
When integrated over the age of the starburst, even for low supernova
rates -- e.g. the current rate within 30\,pc 
of the nucleus of NGC\,3227 is $\sim0.01$\,yr$^{-1}$ \citep{dav06} --
this represents a substantial mass of gas that has been affected by 
supernova remnants.
The STARS model we have constructed in Fig.~\ref{fig:starstoy}
indicates that typically one could expect $\sim10^6$ supernovae to
occur as a result of one of the short-lived starbursts;
and that most of these will occur around 10--50\,Myr after the
beginning of the starburst.
For a decay timescale of the star formation rate that is longer
than $\tau_{\rm SF}=10$\,Myr, this timespan will increase.
Hence, supernovae may also play a role in causing the observed delay
between starburst and AGN activity.

STARS calculates the mass loss and mass loss rates using a very simple scheme,
assuming that a star ejects all of its lost mass at the end of its life on a
stellar track.
Thus, it does not calculate the mass lost from supernovae explicitly, rather
the combined mass lost from OB winds and supernovae which is much higher.
We therefore adopt the $\sim5$\,M$_\odot$ per supernova given above, which
yields a total ejected mass of $\sim8\times10^6$\,M$_\odot$.
This is about 40\% of that released by AGB winds.
However, since this gas is ejected at high speed and 
$\dot{M}_{\rm BH} \propto V_{\rm wind}^{-4}$, the efficiency
with which it can be accreted onto the black hole is extremely low.
This can also be seen in the total mechanical energy of $\sim10^{50}$\,J,
which is several orders of magnitude greater than for AGB winds.
In fact the total mechanical energy exceeds the
binding energy of the nuclear region, which is 
of order $10^{48}$\,J (assuming $10^8$\,M$_\odot$ within 40\,pc).
As a result, it is highly likely that supernova cause some fraction of the gas
to be permanently expelled.
Indeed, superwinds driven by starbursts are well known in many galaxies.
This is not important as long as sufficient gas either
remains to fuel the AGN, or more is produced by stellar winds -- which, as we
have argued above, appears to be the case for AGB stars.

\section{Conclusions}
\label{sec:conc}

We have obtained near infrared spectra of 9 nearby active galactic
nuclei using adaptive optics to achive high spatial resolution (in
several cases better than 10\,pc).
For 7 of these, integral field spectroscopy
with SINFONI allows us to reconstruct the full 2-dimensional
distributions and kinematics of the stars and gas.
Although the individual AGN are very varied, we have analysed them in
a consistent fashion to derive: the stellar K-band luminosity, the
dynamical mass, and the equivalent width of the Br$\gamma$ line. 
We have combined these with radio continuum data from the literature,
which has been used to estimate the supernova rate. 
We have used these diagnostics to constrain STARS evolutionary
synthesis models and hence characterize the star formation timescales
and ages of the starbursts close around AGN.
Our main conclusions can be summarised as follows:

\begin{itemize}
\item
The stellar light profiles show a bright nuclear component with a
half-width at half-maximum of less than 50\,pc.
In a number of cases these nuclear components clearly stand out above
an inward extrapolation of the profile measured on larger scales.
In addition, there are 2 cases which show kinematical evidence for
a distinct stellar component, indicating that the nuclear stellar
populations most probably exist in thick nuclear disks.
The mean mass surface densities of these disks exceeds
$10^4$\,M$_\odot$\,pc$^{-2}$.

\item
There is abundant evidence for recent star formation in the last
10--300\,Myr.
But the starbursts are no longer active, implying that the star
formation timescale is short, of order a few tens of Myr.
While the starbursts were active, the star formation rates would have
been much higher than the current rates, reaching as high as
1000\,M$_\odot$\,kpc$^{-2}$ in the central few tens of parsecs
(comparable to ULIRGs, but on smaller spatial scales).
These starbursts would have been Eddington limited.
Due to the very high star forming efficiency, the starbursts would
have also exhausted their fuel supply on a short timescale and hence
have been short-lived.
It therefore seems likely that nuclear starbursts are episodic in
nature.

\item
There appears to be a delay of 50--100\,Myr (and in some cases perhaps more)
between the onset of star formation and the onset of AGN activity.
We have interpreted this as indicating that the starburst has a significant
impact on fuelling the central black hole, and
have considered whether outflows from stars might be responsible.
While supernovae and winds from OB stars eject a large mass of gas, the high
velocity of this gas means that its accretion efficiency is extremely low.
On the other hand, winds from AGB stars ultimately dominate the total mass
ejected in a starburst;
and the very slow velocities of these winds mean they can be accreted onto the
black hole very efficiently.

\end{itemize}


\acknowledgments

The authors thank all those who assisted in the observations, and also
the referee for a thorough review of the paper.
This work was started at the Kavli Institute for Theoretical Physics
at Santa Barabara and as a result was supported in part by the
National Science Foundation under Grant No. PHY05-51164. 
RD aknowledges the interesting and useful discussions he had there
with Eliot Quartaert, Norm Murray, Julian Krolik and Todd Thompson.

{\it Facilities:} 
\facility{Keck:II (NIRSPAO, NIRC2)}, 
\facility{VLT:Yepun (NACO, SINFONI)}.

\appendix

\section{Individual Objects}
\label{sec:obj}

This appendix contains specific details on the individual objects.
We summarize our published results from near infrared
adaptive optics spectroscopy of individual objects, and present a
brief analysis of the new data for several other objects.
The aim of re-assessing the data for Mkn\,231 that has already been
published is to ensure that it is analysed using STARS in a manner
that is consistent with the new data.
For NGC\,7469, we make a significant update of the analysis using new
data from integral field spectroscopy.
In general, for objects with new data, we provide only the part of the
analysis relevant to understanding star formation around the AGN.
Our intention is that a complete analysis for each object will be
presented in future publications.

Our analyses are restricted to the nuclear region.
Since there is no strict universal definition of what comprises the
`nuclear region', we explicitly state in Table~\ref{tab:derprop} the
size of the region we study in each galaxy.
The table also presents a summary of the primary diagnostics.
The way in which these have been derived, and their likely
uncertainties, has been discussed in some detail already in
Section~\ref{sec:diag}.
As such, the description of these methods is not repeated, and in this
Section we discuss only issues that require special attention.

\subsection{Summary of Star Forming Properties of Galaxies already Studied}

\subsubsection{Mkn 231}

A detailed analysis of the star formation in the nucleus of Mkn\,231 at a
resolution of about 0.18\arcsec\ (150\,pc) was given in \cite{dav04b}.
Here we summarize only the main points; no new data is presented, but the
analysis is updated using STARS to make it consistent with the other
objects studied in this paper.

The presence of stellar absorption features across the nucleus demonstrates the
existence of a significant population of stars.
The radial distribution and kinematics indicate they lie, like the gas
\citep{dow98}, in a nearly face-on disk.
\cite{dav04b} found that the dynamical mass imposed a strong
constraint on the range of acceptable starburst models, yielding 
an upper limit to the age of the stars of around 120\,Myr.
Re-assessing the mass-to-light ratio using STARS models suggests that
for the increased mass required by a more face-on orientation ($i=10^\circ$)
an upper age of 250\,Myr is also possible, depending on the star formation
timescale. 
However, either a small change of only a few degrees to the
inclination (to $i=15^\circ$),
or a relatively short star formation timescale of 10\,Myr would reduce
the limit to the $\sim$100\,Myr previously estimated.
This is more consistent with the extremely high supernova rate.

The stellar luminosity, found from the dilution of the CO absorption
\citep{dav04b}, indicates that stars within 1\arcsec\ (800\,pc) of the nucleus
contribute 25--40\% of the bolometric luminosity of the galaxy.
Similarly, within 200\,pc, stars comprise 10--15\% of L$_{\rm bol}$.
The age, star formation rate, and size scale (disk scale length of
0.18--0.2\arcsec) are all consistent with high resolution radio continuum
imaging \citep{car98}.

\subsubsection{Circinus}

Star formation in the central 16\,pc of Circinus was addressed by
\cite{mul06}.
The diagnostics given in Table~\ref{tab:derprop} are taken
from this reference.
We used the depth of the CO\,2-0 bandhead to estimate the stellar
luminosity, combined with the narrow Br$\gamma$ flux (which we argued
originated in star forming regions rather than the AGN narrow line
region) and the radio continuum, to constrain starburst models.
The conclusion was that the starburst was less than 80\,Myr old and
was already decaying.
On these scales it contributes 1.4\% of L$_{\rm bol}$, or more if
extinction is considered.
A similar nuclear star formation intensity was estimated by
\cite{mai98}, who were also able to study Circinus on larger scales.
They found that the luminosity of young stars within 200\,pc of
the AGN was of order $10^{10}$\,$L_\odot$, and hence comparable to the AGN.

\subsubsection{NGC 3227}

An analysis similar to that for Circinus was performed on NGC\,3227 by
\cite{dav06}, and the diagnostics given in Table~\ref{tab:derprop} are taken
from this reference.
In this case we were able to make estimates of and correct for
contributions of:
(1) the narrow line region to Br$\gamma$, because
there were clear regions along the minor axis that had higher
dispersion; 
(2) the AGN to the radio continuum, by estimating the maximum contribution
from an unresolved source; and 
(3) the bulge stars to the stellar luminosity, by extrapolating the
radial profile of the bulge to the inner regions.
The STARS models yielded the result that in the nucleus,
star formation began approximately 40\,Myr ago and must have already ceased.
At the resolution of 0.085\arcsec, the most compact component of stellar
continuum had a measured FWHM of 0.17\arcsec, suggesting an intrinsic size
scale of $\sim12$\,pc.
Young stars within 30\,pc of the AGN (i.e. more than just the most compact
region) have a luminosity of $\sim3\times10^{9}$\,$L_\odot$,
which is $\sim20$\% of the entire galaxy.

\subsection{Star Forming Properties of Galaxies with New Data}

\subsubsection{NGC 7469}

Star formation on large scales in NGC\,7469 has been studied by
\cite{gen95}.
They found that within 800\,pc of the nucleus, a region that includes the
circumnuclear ring, the luminosity from young stars was
$\sim3\times10^{11}$\,$L_\odot$, about 70\% of the galaxy's bolometric
luminosity. 
This situation is similar to that in Mkn\,231.
On smaller scales, the nuclear star formation in NGC\,7469 was directly
resolved by \cite{dav04a} 
on a size scale of 0.15--0.20\arcsec\ (50--65\,pc) FWHM.
An analysis of the longslit data, similar to that for Mkn\,231, was made --
making use of 
stellar absorption features, kinematics, and starburst models.
We estimated that the age of this
region was no more than 60\,Myr under the assumption that the fraction of
stellar light in the K-band in the central 0.2\arcsec\ was 20--30\%.
Our new integral field SINFONI observations of NGC\,7469 at a spatial
resolution of 0.15\arcsec\ (measured from both the broad Br$\gamma$
and the non-stellar continuum profiles, see Section~\ref{sec:obs})
are used here to make a more accurate estimate of the nuclear
K-band luminosity. 
They enable us to provide a short update to the detailed analysis in
\cite{dav04a}.

The SINFONI data show that the equivalent width of the
2.3\micron\ CO\,2-0 is $W_{\rm CO\,2-0} = 1.8$\AA\ in a 0.8\arcsec\ aperture
and 0.9\AA\ in a 0.2\arcsec\ aperture.
The corresponding K-band magnitudes are $K=10.4$ and $K=11.8$ respectively.
If one takes the intrinsic equivalent width of the 2.3\micron\ CO\,2-0
bandhead to be 12\AA\ (see Section~\ref{sec:diag}), one arrives at a
more modest value 
of 8\% for the stellar fraction of K-band continuum in the 0.2\arcsec\
aperture.
The stellar K-band luminosity in this region is then $6\times10^7$\,L$_\odot$.
Comparing this to the dynamical mass in \cite{dav04a} yields a mass-to-light
ratio of M/L$_K \sim 0.6$\,M$_\odot$/L$_\odot$.
Previously, extrapolation from a 37\,mas slit to a filled aperture had
led to an underestimation of the total magnitude but an overestimation
of the stellar contribution.
Fortuitously, these uncertainties had compensated each other.
The same analysis for the 0.8\arcsec\ aperture yields a K-band stellar
luminosity of $3\times10^8$\,L$_\odot$ and hence 
M/L$_K \sim 1.6$\,M$_\odot$/L$_\odot$.

The K-band datacube yields estimates of the upper limit to 
$W_{\rm Br\gamma}$ of 17\AA\ and 11\AA\ in 0.2\arcsec\ and 0.8\arcsec\
apertures respectively.
This has been corrected for dilution of the stellar continuum (as described in
Section~\ref{sec:diag}) but not
for a possible contribution to the narrow Br$\gamma$ from the AGN.
Hence the actual $W_{\rm Br\gamma}$ corresponding to only the stellar
line and continuum emission will be less than these values --
indicating that the star formation is unlikely still to be on-going.

We estimate the age of the star formation using the STARS models in
Fig.~\ref{fig:stars}.
Within the 0.2\arcsec\ aperture this gives 100\,Myr, comparable to our
original estimate.
Such a young age is supported by radio continuum measurements.
With a 0.2\arcsec\ beam, \cite{col01} reported that the unresolved
core flux in NGC\,7469 was 12\,mJy at 8.4\,GHz.
With much higher spatial resolution of 0.03\arcsec, \cite{sad95}
reported an upper limit to the unresolved 8.4\,GHz continuum of
7\,mJy.
We assume that the difference of 5\,mJy is due to emission extended on scales
of 10--60\,pc which is resolved out of one beam but not the other.
As discussed in Section~\ref{sec:diag}, star formation is a likely candidate
for such emission.
In this case, we would estimate the supernova rate to be
$\sim0.1$\,yr$^{-1}$ and the ratio 
$10^{10}\,\nu_{\rm SN}/L_K \sim 3$.
This is likely to be a lower limit since there was only an upper limit
on the core radio flux density.
For a ratio of this order, even allowing for some uncertainty,
Fig~\ref{fig:stars} implies an age 
consistent with no more than 100\,Myr.

Within the 0.8\arcsec\ aperture, which we adopt in
Table~\ref{tab:derprop}, continuous star formation is inconsistent
with W$_{\rm Br\gamma}$.
For a star formation timescale of $\tau_{\rm SF}=100$\,Myr, the
mass-to-light ratio
implies an age of 190\,Myr, just consistent with the measured value of 
W$_{\rm Br\gamma}=11$\AA. 
If some of the narrow Br$\gamma$ is associated with the AGN rather
than star formation, then a shorter star formation timescale is
required.
For $\tau_{\rm SF}=10$\,Myr, the ratio M/L$_K$ yields an age of
110\,Myr.

\subsubsection{IRAS 05189-2524}

Fig.~\ref{fig:ir05189_spec} shows the H-band spectrum integrated across two
segments of the NACO slit, located on either side of the nucleus.
It shows that even away from the nucleus, the depth of the stellar
absorption features is only a few percent.
We have therefore decomposed the data into the stellar and non-stellar
parts using both the stellar absorption features and the spectral
slope of the continuum.
The latter method has been shown to work for well sampled data by
\cite{dav04a}.
The rationale is that the hot dust associated with the AGN will be
much redder than the stellar continuum.
An AGN component is also expected to be unresolved for a
galaxy at the distance (170\,Mpc) of IRAS\,05189-2524.
The spectral slope was determined by fitting a linear function to the
spectrum at each spatial position along the slit.
It is plotted as a function of position in
Fig.~\ref{fig:ir05189_decomp}, showing a single narrow peak.
A Gaussian fit to this yields a spatial resolution of 0.12\arcsec\ (100\,pc)
FWHM.
The stellar continuum, also shown in Fig.~\ref{fig:ir05189_decomp},
has been determined by summing the four most prominent 
absorption features: CO\,4-1, Si\,I, CO\,5-2, CO\,6-3.
While a Gaussian is not an optimal fit to this profile, it does yield an
aproximate size scale, which we find to be 0.27\arcsec\ FWHM.
Quadrature correction with the spatial resolution yields an intrinsic size of
0.25\arcsec\ (200\,pc).
As a cross-check, in the figure we have compared the sum of these two
components to the full continuum profile.
The good match indicates that the decomposition appears to be reasonable.

Remarkably, the 200\,pc size of the nuclear stellar light is very similar to
that of the 8.44\,GHz radio continuum map of \cite{con91}.
With a beam size of 0.50\arcsec$\times$0.25\arcsec, they resolved the nuclear
component to have an intrinsic size of 0.20\arcsec$\times$0.17\arcsec.
In constrast to radio sources which are powered by AGN and have
brightness temperatures $T_{\rm b}\gg10^5$\,K, 
the emission here is resolved and has a low brightness
temperature of $\sim4000$\,K.
This implies a star forming origin.
Using their scaling relations further suggests that the flux density
corresponds to a supernova rate of $\sim1$\,yr$^{-1}$.

As described in Section~\ref{sec:diag}, we have estimated the stellar
luminosity by comparing the H-band 
spectrum to a template star to correct for dilution.
We used HR\,8465 a K\,1.5\,I star for which the equivalent width of
CO\,6-3 is 4.2\AA, within the 4--5\AA\ range predicted by STARS in
Fig.~\ref{fig:equiv}. 
By extrapolating from the spatial profiles along the slit we have estimated
the integrated equivalent width within a 1.1\arcsec\ aperture, for which
\cite{sco00} gave an H-band magnitude of 11.83.
Using all four features above we find for the template $W=14.4$\AA\ and for
IRAS\,05189-2524 $W_{\rm CO6-3}=6.7$\AA.
This implies that in the central 1.1\arcsec, approximately 45\% of the H-band
continuum originates in stars.
Using the colour conversion $H-K=0.15$ from Fig.~\ref{fig:col_lum}
(see Section~\ref{sec:diag}) we find a K-band magnitude for the 
stars of 12.55\,mag and hence a K-band stellar luminosity of
$2\times10^9$\,L$_\odot$.
Putting these results together we derive a ratio of supernova rate to K-band
stellar luminosity of 
$\nu_{\rm SN} [yr^{-1}] / L_{K} [10^{10} L_\odot] \sim 5$.
Applying corrections for extinction and an AGN contribution would tend to
decrease this ratio.

As a second diagnostic we use $W_{\rm Br\gamma}$.
We estimate the dilution of the K-band continuum via two
methods.
Firstly, we measure $W_{\rm Na I} = 0.3$\AA, indicating a stellar
fraction of 0.10--0.15.
A consistency check is provided by the H-band dilution, which we
extrapolate to the K-band using blackbody functions for the
stars and dust assuming characteristic temperatures of 5000\,K and
1000\,K respectively.
This method suggests the K-band stellar fraction is around $\sim0.14$.
Hence correcting the directly measured equivalent width of the narrow
Br$\gamma$ for the non-stellar continuum yields 
$W_{\rm Br\gamma}=4$--5\AA.

Since IRAS\,05189-2524 is close to face-on \citep{sco00}, it is not
straightforward to make a reliable estimate of the dynamical mass.
Nevertheless, requiring $\nu_{\rm SN} / L_{K} $ to be high while 
$W_{\rm Br\gamma}$ is low already puts significant constraints on the
star formation history.
Thus, although the star formation has probably ended, the age is
unlikely to be greater than 100\,Myr, and could be 
as low as 50\,Myr where $\nu_{\rm SN} / L_{K} $ peaks.
For such ages the ratio $L_{\rm bol}/L_K$ is in the range 100--150. 
Hence for the young stars within 0.55\arcsec\ (450\,pc) of the nucleus we find
$L_{\rm bol} \sim ($2--3$) \times 10^{11}$\,L$_\odot$, about 20\% of 
$L_{\rm bol}$ for the galaxy.

\subsubsection{NGC 2992}

The spatial resolution of the K-band data for NGC\,2992 has been estimated
from both the broad Br$\gamma$ and the non-stellar continuum (see
Section~\ref{sec:obs} and~\ref{sec:diag}).
The two methods yield symmetric PSFs, with FWHMs of 0.32\arcsec\ and
0.29\arcsec\ respectively, corresponding to 50\,pc.

The CO\,2-0 equivalent width of $\sim3$\AA\ implies a stellar fraction of
$\sim0.25$ within a radius of 0.4\arcsec, and hence a stellar
luminosity of L$_K = 3.5\times10^7$\,L$_\odot$.

Unlike IRAS\,05189-2524, the radio continuum in NGC\,2992 is quite complex.
Much of the extended emission on scales of a few arcsec appears to originate
from a superbubble, driven either by the AGN or by a nuclear starburst.
On the other hand, most of the nuclear emission seems to be unresolved.
With a beam size of 0.34\arcsec$\times$0.49\arcsec, \cite{whe88} measured the
unresolved flux to be 7\,mJy at 5\,GHz. 
At a resolution better than 0.1\arcsec, \cite{sad95} reported a 2.3\,GHz
flux of 6\,mJy.
Based on this as well as non-detections at 1.7\,GHz and 8.4\,GHz, they
estimated the core flux at 5\,GHz to be $<$6\,mJy.
Taking a flat spectral index, as indicated by archival data \citep{cha00}, one
might expect the 5\,GHz core flux to be not much less than 6\,mJy, 
leaving room for only $\sim1$\,mJy in extended emission in the central
0.5\arcsec.
If we assume this difference can be attributed to star formation, it
implies a supernova rate of $\sim0.003$\,yr$^{-1}$ and hence  
$10^{10}\,\nu_{\rm SN} / L_{K} \sim 1$.
Fig.~\ref{fig:stars} shows that a ratio of this order is what one
might expect for ages up to 200\,Myr.
However, given the uncertainty it does not impose a significant
constraint.

It is also difficult to quantify what fraction of the narrow
Br$\gamma$ is associated with star formation.
This is made clear in Fig.~\ref{fig:n2992_brg} which shows that the morphology
of the line (centre left panel) does not follow that of the stars (far left).
In addition, particularly the south-west side is associated with velocities
that are bluer than the surrounding emission, indicative of motion towards us.
The western edge also exhibits high velocity dispersion.
Taken together, these suggest that we may be seeing outflow from the apex of an
ionisation cone with a relatively large opening angle.
This interpretation would tend to support the hypothesis that the
radio bubble has been driven by the AGN.

The stellar continuum appears to trace an inclined disk, the north
west side of which is more obscured (Fig.~\ref{fig:n2992_brg}).
However, the velocity dispersion is high, exceeding 150\,km\,s$^{-1}$
across the whole field (Fig.~\ref{fig:n2992_vel}). 
This is similar to the 160\,km\,s$^{-1}$ reported by \cite{nel95}
from optical spectroscopy, and suggests that we are seeing bulge stars.
To analyse the radial luminosity profile we have fitted it with both
an $r^{1/4}$ law and exponential profile.
The fits in Fig.~\ref{fig:n2992_rad} were optimised at radii
$r>0.5$\arcsec\ and then extrapolated inwards, convolved with the PSF.
Whether one could claim that there is excess continuum in the
nucleus depends on the profile fitted.
The $r^{1/4}$ law provides a
stronger constraint since it is more cuspy, and suggests there is no
excess.
Although this evidence is inconclusive, 
Fig.~\ref{fig:n2992_vel} suggests that there is some kinematic evidence
favouring the existence of a distinct 
nuclear stellar population.
This comes in the form of a small unresolved drop in
dispersion at the centre, similar to those in NGC\,1097 and NGC\,1068.
While the evidence in NGC\,2992 is not compelling, the dispersion is
consistent with there being an equivalent -- but fainter -- nuclear disk
on a scale of less than our resolution of 50\,pc.
In general it seems that the K-band light we are seeing is dominated by
the bulge, and we are therefore unable to probe in detail the inner
region where it seems that more recent star formation has probably
occurred.

Thus, although the available data suggest there has likely been
recent star 
formation in the nucleus of NGC\,2992, the only strong constraint we can
apply is that continuous star formation in the central arcsec over the
last Gyr can be ruled out since it would 
require $W_{\rm Br\gamma} > 10$--15\AA.
We therefore omit NGC\,2992 from the discussion and analysis in
Sections~\ref{sec:prop} and~\ref{sec:starAGN}.

\subsubsection{NGC 1097}

In NGC\,1097, the first evidence for recent star formation near the nucleus
was in the form of a reduction in the stellar velocity dispersion.
\cite{ems01} proposed this could be explained by the presence of a
dynamically cold nuclear disk that had recently formed stars.
Direct observations of a spiral structure in the central few arcsec, from
K-band imaging \citep{pri05} and [N\,{\sc ii}] streaming motions \citep{fat06},
have since confirmed this idea.
However, some issues remain open, such as
why there are three spiral arms rather than the usual two, and why gas along
one of them appears to be outflowing.

Our data, at a resolution of 0.25\arcsec\ measured from the H-band
non-stellar continuum, also reveal the same spiral structure.
Indeed, we find that it is traced by the morphology of the CO bandhead
absorption as well as by the 2.12\,\micron\ H$_2$ 1-0\,S(1) line.
Interestingly, 1-0\,S(1) emission is stronger where the stellar features are
weaker.
This suggests that obscuration by gas and dust plays an important role.
Fig.~\ref{fig:n1097_prof} shows that an $r^{1/4}$ law, typical of
stellar bulges, with effective radius $R_{\rm eff} = 0.5$\arcsec\ is a
good fit to the stellar radial profile at $0.5\arcsec<r<1.8$\arcsec.
It therefore seems reasonable to argue that at these radii it is only
the gas that lies in a disk. 
In this picture the spiral structure in the stellar continuum arises
solely due to extinction of the stars behind the disk.
Extrapolating this fit, convolved with the PSF, to the nucleus indicates that
at $r<0.5$\arcsec\ there is at least 25\% excess stellar continuum.
There could be much more, given that it coincides with a change in
the dominant kinematics.

For NGC\,1097 we parameterized the kinematics of the gas and stars
quantitatively using kinemetry.
Based on the uniformity of the velocity field, we made the simplifying
assumption that across the central 4\arcsec\ the gas lies in a single
plane whose centre is coincident with the peak of the non-stellar
emission.
We were then able to derive the position angle and inclination of the
disk (see Section~\ref{sec:obs}).
The 2D kinematics of the stars is traced via the CO2-0 absorption
bandhead, and that of the gas through the 1-0S(1) emission line.
These independently yielded similar
parameters: both gave a position angle of $-49^\circ$ and their
inclinations were 43$^\circ$ and 32$^\circ$ respectively.
These are fully consistent
with values found by other authors (\citealt{sto03,fat06}).
The resulting rotation curves and velocity dispersions are shown in
Fig.~\ref{fig:n1097_dispvel}.
The residuals, which can be seen in the velocity field of the gas but
not the stars, and their relation to the spiral structure described
above will be discussed elsewhere (Davies et al. in prep).
The important result here is that at our spatial resolution, we find
that the central stellar
dispersion is $\sigma_*=100$\,km\,s$^{-1}$, less than the
surrounding 150\,km\,s$^{-1}$ and also less than that in the
seeing limited spectra of \cite{ems01}.
In the same region we find that the rotation velocity of the gas starts to
decrease rapidly, and its dispersion increases from 
$\sigma_{\rm gas}\sim40$\,km\,s$^{-1}$ to $\sim80$\,km\,s$^{-1}$.

Fig.~\ref{fig:n1097_dispvel} also shows that while the
kinematics of the stars and gas are rather different at large
($>0.5$\arcsec) radii, they are remarkably similar at radii $<0.5$\arcsec.
This certainly provides a strong indication that in the nuclear
region the stars and gas are coupled, most likely in a (perhaps thick)
disk;
and that the stars in this disk, which are bright and hence presumably
young, give rise to the excess stellar continuum observed.

Evidence for a recent starburst has been found by \cite{sto05} through
optical and UV spectra.
They argued that a number of features they observed could only arise from an
$10^6$\,M$_\odot$ instantaneous starburst, which occurred a few Myr ago
and is reddened by $A_V=3$\,mag of extinction.
Using STARS we have modeled this starburst as
a $10^6$\,M$_\odot$ burst beginning 8\,Myr ago with an exponential
decay timescale of 1\,Myr.
The age we have used is a little older to keep the Br$\gamma$
equivalent width low; and at this age, the model predicts 
$W_{\rm Br\gamma}=4$\AA.
As Fig.~\ref{fig:n1097_maps} shows, the observed Br$\gamma$ is weak,
although perhaps slightly resolved.
Corrected for the non-stellar continuum, we measure only 
$W_{\rm Br\gamma} \sim 1$\AA.
However, the bulge population may account for a
significant fraction of the K-band stellar continuum.
Correcting also for this could increase
$W_{\rm Br\gamma}$ to 2--5\AA, consistent with that
of the model -- assuming that the Br$\gamma$ is associated with the starburst
rather than the AGN.
To within a factor of a few, the scale of the model starburst is also
consistent with that measured:
In the central 0.5\arcsec\ we measure a Br$\gamma$ flux of
$2\times10^{-19}$\,W\,m$^{-2}$, compared to that predicted by the model of 
$5\times10^{-19}$\,W\,m$^{-2}$.
Given the uncertainties -- factors of a few -- both in the parameters
of the starburst model and also
in the corrections we have applied to the data, we consider this
a good agreement.

We cannot constrain the starburst further due to its compactness.
\cite{sto05} found that it was occurring in the central 0.2\arcsec, 
whereas our resolution is only 0.25\arcsec.
The Br$\gamma$ emission is confined to the central 0.4--0.5\arcsec,
although its size is hard to measure due to its weakness with respect
to the stellar absorption features.
In this region the K-band stellar luminosity is
$4.5\times10^6$\,L$_\odot$.
To estimate the dynamical mass we use the mean kinematics of the stars
and gas, i.e. V$_{\rm rot}=40$\,km\,s$^{-1}$ (corrected for inclination) and 
$\sigma=90$\,km\,s$^{-1}$ (this is the central value, which is least
biassed by bulge stars), yielding 
$1.4\times10^8$\,M$_\odot$.
This is actually dominated by the black hole, which has a mass of 
$(1.2\pm2)\times10^8$\,M$_\odot$ \citep{lew06}.
The difference between these implies a mass of gas and stars of 
$\sim2\times10^7$\,M$_\odot$, although with a large uncertainty.
The associated mass-to-light ratio is $M/L_K\sim4$.
On its own, this implies that 
over the relatively large area that it encompasses, 
the maximum characteristic age for the star formation is a few hundred Myr.
If one speculates that star formation has been occurring sporadically
for this timescale, then the starburst seen by \cite{sto05} is the
most recent active episode.

In order to make a rough estimate of the supernova rate in the central
region we make use of measurements reported by \cite{hum87}.
They find an unresolved component (size $<0.1$\arcsec) with 5\,GHz flux
density $3.5\pm0.3$\,mJy, but at lower resolution there is a
$4.1\pm0.3$\,mJy component of size 1\arcsec.
As discussed in Section~\ref{sec:diag} we assume that the difference
-- albeit with only marginal significance -- of $0.6\pm0.4$\,mJy 
is due to star formation in the central region, which implies a
supernova rate of $6\times10^{-4}$\,yr$^{-1}$ and hence 
$10^{10}\,\nu_{\rm SN}/L_K \sim 1.3$, a value consistent with rather
more recent star formation.
Indeed, when compared to Fig.~\ref{fig:stars}, this and the low
$W_{\rm Br\gamma}$ imply a young age and short star formation
timescale.
For $\tau_{\rm SF}=10$\,Myr the age is 60--70\,Myr;
for an instantaneous burst of star formation, the age would be
$\sim10$\,Myr, broadly consistent with that of \cite{sto05}.

Thus, although our data do not uniquely constrain the age of the
starburst in the nucleus of NGC\,1097, they do indicate that recent
star formation has occurred; and they are consistent with
a very young compact starburst similar to that derived from optical
and UV data.

\subsubsection{NGC 1068}

Evidence for a stellar core in NGC\,1068 with an intrinsic size scale of
$\sim45$\,pc was first presented by \cite{tha97}.
Based on kinematics measured in large (2--4\arcsec) apertures, they assumed
the core was virialized and estimated a mass-to-light ratio based on this
assumption leading to an upper limit on the stellar age of 1600\,Myr.
Making a reasonable correction for an assumed old component lead to a younger
age of 500\,Myr.

Stellar kinematics from optical integral field spectra \citep{ems06a,ger06}
show evidence for a drop in the stellar velocity dispersion in the central few
arcsec to $\sigma_* \sim 100$\,km\,s$^{-1}$, inside a
region of higher 150--200\,km\,s$^{-1}$ dispersion (presumably the bulge).
Our near infrared adaptive optics data are able to fully resolve the inner
region where $\sigma_*$ drops, as shown in Fig.~\ref{fig:n1068_prof}.
As for NGC\,1097, the velocity distribution of the stars was derived
through kinemetry, again making use of the uniformity of
the stellar velocity field to justify the simplifying assumption that
the position 
angle and inclination do not change significantly in the central
4\arcsec.
The derived inclination of $40^\circ$ and position angle of
$85^\circ$ are quantitatively similar to those found by other authors
in the central few to tens of arcseconds \citep{ems06a,ger06,gar99}.
The uniformity of the stellar kinematics is in contrast to molecular
gas kinematics, as traced via the 1-0\,S(1) line, which are strongly perturbed
and show several distinct structures superimposed.
These are too complex to permit a comparably simple analysis and will
be discussed, together with the residuals in the stellar kinematics
in a future work (Mueller S\'anchez et al. in prep).

The crucial result relevant here is that 
at our H-band resolution of 0.10\arcsec\, we find that $\sigma_*$
reduces from 130\,km\,s$^{-1}$ at 1--2\arcsec\ to only
70\,km\,s$^{-1}$ in the very centre.
That there is in the same region an excess in the stellar continuum is
demonstrated in Fig.~\ref{fig:n1068_prof_wide}.
Here we show the radial profile of the stellar continuum from both
SINFONI integral field spectra out to a radius of 2\arcsec\ and NACO
longslit spectra out to 5\arcsec\ (350\,pc).
At radii 1--5\arcsec, corresponding roughly to the region of high stellar
dispersion measured by \cite{ems06a}, 
the profile is well matched by an $r^{1/4}$ law, as
one might expect for a bulge.
At radii $r<1$\arcsec\ -- the same radius at which we begin to see a
discernable reduction in the stellar dispersion -- the stellar continuum
increases by as much as a factor 2 above the inward extrapolation of
the profile, indicating that there 
is extra emission.
As for NGC\,1097, the combined signature of dynamically cool
kinematics and excess emission is strong evidence for a nuclear disk which has
experienced recent star formation.

We can make an estimate of the characteristic age of the star formation in
the central arcsec based on the mass-to-light ratio in a similar way to
\cite{tha97}. 
Because the stars appear to lie in a disk, we estimate the dynamical
mass as described in Section~\ref{sec:diag} from the stellar 
kinematics, using the rotation velocity and applying a correction for
the dispersion.
The stellar rotation curve is essentially flat at 0.1--0.5\arcsec, with
$V_* = 45$\,km\,s$^{-1}$ (corrected for inclination).
We also take $\sigma_* = 70$\,km\,s$^{-1}$, which is the central value
and hence least biassed by the high dispersion bulge stars.
These lead to a mass of $1.3\times10^8$\,M$_\odot$ within 
$r=0.5$\arcsec\ (35\,pc), and 
a mean surface density of $3\times10^4$\,M$_\odot$\,pc$^{-2}$.
Correcting for the non-stellar continuum, the H-band magnitude (which the
behaviour of $\sigma_*$ indicates is dominated by the disk emission) in the
same region is 11.53\,mag.
For $H-K=0.15$\,mag (Fig.~\ref{fig:stars}), we find
$L_K=4.3\times10^7$\,L$_\odot$ and hence
$M/L_K = 3$\,M$_\odot$/L$_\odot$.
If no star formation is on-going, this implies a characteristic age of
200--300\,Myr fairly independent of the timescale (for 
$\tau_{\rm SF} \lesssim 100$\,Myr, see Fig.~\ref{fig:stars}) on which
stars were formed.
We note that this is significantly younger than the age estimated by
\cite{tha97} primarily because their mass was derived using a higher
$\sigma_*$ corresponding to the bulge stars.

The assumption of no current star formation is clearly demonstrated by
the Br$\gamma$ map in Fig.~\ref{fig:n1068_maps}.
Away from the knots of Br$\gamma$, which are associated with the coronal lines
and the jet rather than possible star formation, the equivalent width is
$W_{\rm Br\gamma}\sim4$\AA. 
This is significantly less than that for continuous star formation of any age.
Thus, while it seems likely that star formation has occurred in the last few
hundred Myr, it also seems an unavoidable conclusion that there is no current
star formation.

To complete our set of diagnostics for NGC\,1068, we consider also the
radio continuum.
This is clearly dominated by phenomena associated with the AGN and
jets, and our best estimate of the flux density away from these
features is given by the lowest contour in maps such as Figure~1 of
\cite{gal04}.
From this we estimate an upper limit to the 5\,GHz continuum
associated with star formation of 128\,mJy within $r<0.5$\arcsec.
However, converting to a supernova rate and comparing to the K-band
stellar luminosity yields a limit that is not useful, being an order
of magnitude above the largest expected values.

\subsubsection{NGC 3783}
\label{sec:ngc3783}

At near infrared wavelengths, the AGN in NGC\,3783 is remarkably bright.
Integrated over the central 0.5\arcsec\ less than 4\% of the K-band
continuum is stellar.
In addition, the broad Brackett lines are very strong and dominate the
H-band.
Both of these phenomena are immediately clear from the H- and K-band
spectra in Fig.~\ref{fig:n3783_spec}.
However, it does mean that the spatial resolution can be measured
easily from both the non-stellar continuum and the broad emission
lines (see Section~\ref{sec:obs}).
We find the K-band PSF to be symmetrical with a FWHM of 0.17\arcsec.

Due to the ubiquitous Brackett emission in the H-band we were unable
to reliably trace the stellar absorption features and map out
the stellar continuum.
Instead we have used the CO\,2-0 bandhead at 2.3\micron\ even though
the dilution at the nucleus itself is extreme.
The azimuthally averaged radial profile is shown in
Fig.~\ref{fig:n3783_rad} together with the PSF for reference.
At radii from 0.2\arcsec--1.6\arcsec\ (the maximum we can measure) the
profile is well fit by an $r^{1/4}$ de Vaucouleurs law with 
$R_{\rm eff}=0.6$\arcsec\ (120\,pc).
As has been the case previously, at smaller radii we find an excess
that here is perhaps marginally resolved. 
Thus a substantial fraction of the near infrared stellar continuum in
the central region is likely to originate in a population of stars
distinct from the bulge.

We were unable to measure the stellar kinematics due to the limited
signal-to-noise. 
Instead, we used the molecular gas kinematics to estimate the
dynamical mass.
As before, we used kinemetry to derive the position
angle of $-14^\circ$ and the inclination in the range 35--39$^\circ$.
This orientation is consistent with the larger (20\arcsec) scale
isophotes in the J-band 2MASS image and implies that in NGC\,3783
there is no significant warp on scales of 50\,pc to 4\,kpc.
A small inclination is also consistent with its classification as a
Seyfert~1.
Adopting these values, the resulting rotation curve is shown in
Fig.~\ref{fig:n3783_rot}.
At very small radii the rising rotation curve may be the result of
beam smearing across the nucleus.
At $r>0.2$\arcsec, the falling curve suggests that the rotation is
dominated by the central ($r<0.2$\arcsec) mass, perhaps the
supermassive black hole.
We estimate the dynamical mass within a radius of 0.3\arcsec\ (60\,pc),
corresponding to the point where the excess continuum begins and also
where the rotation curve appears to be unaffected by beam smearing.

\begin{figure}
\epsscale{0.32}
\plotone{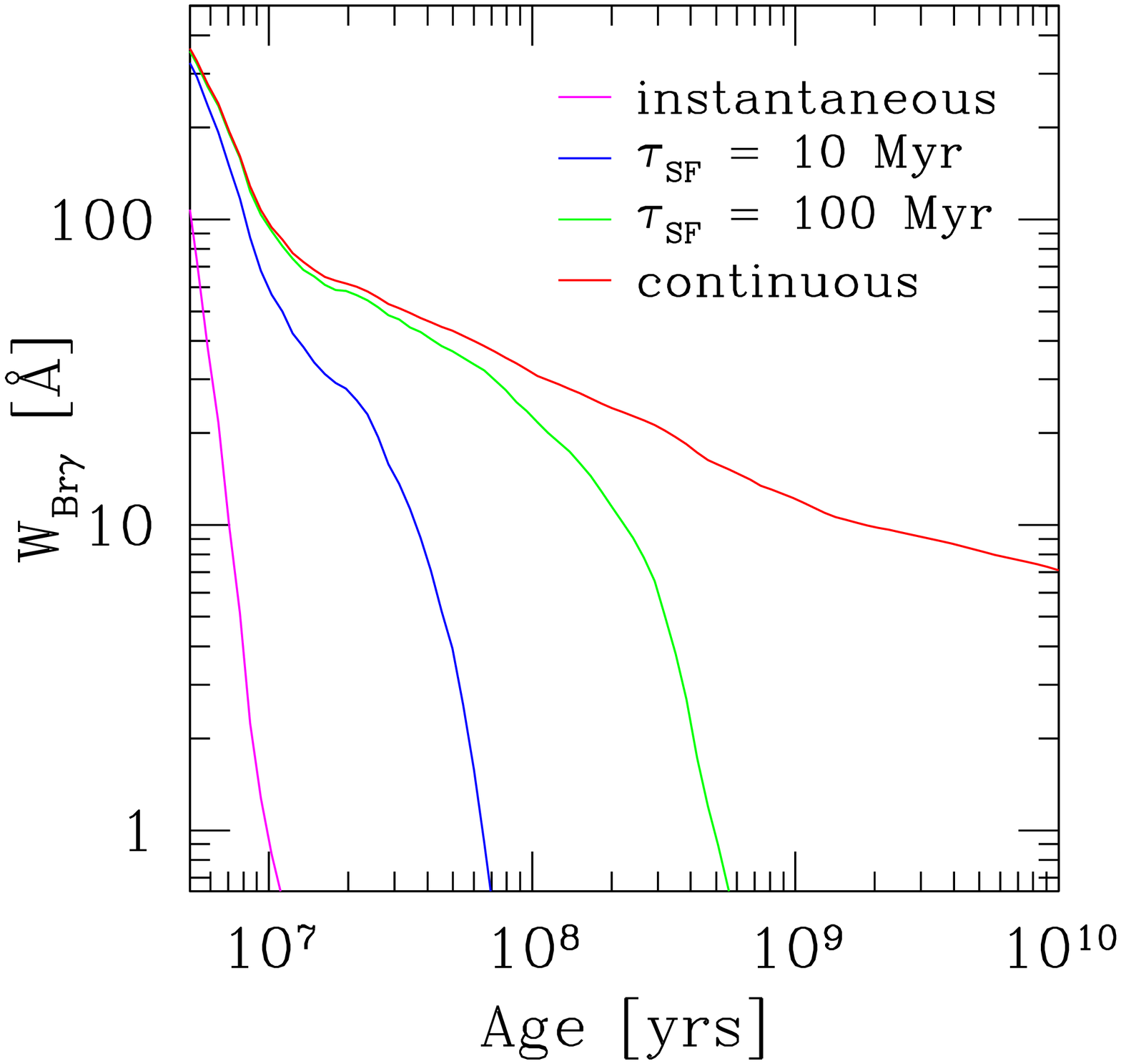} 
\plotone{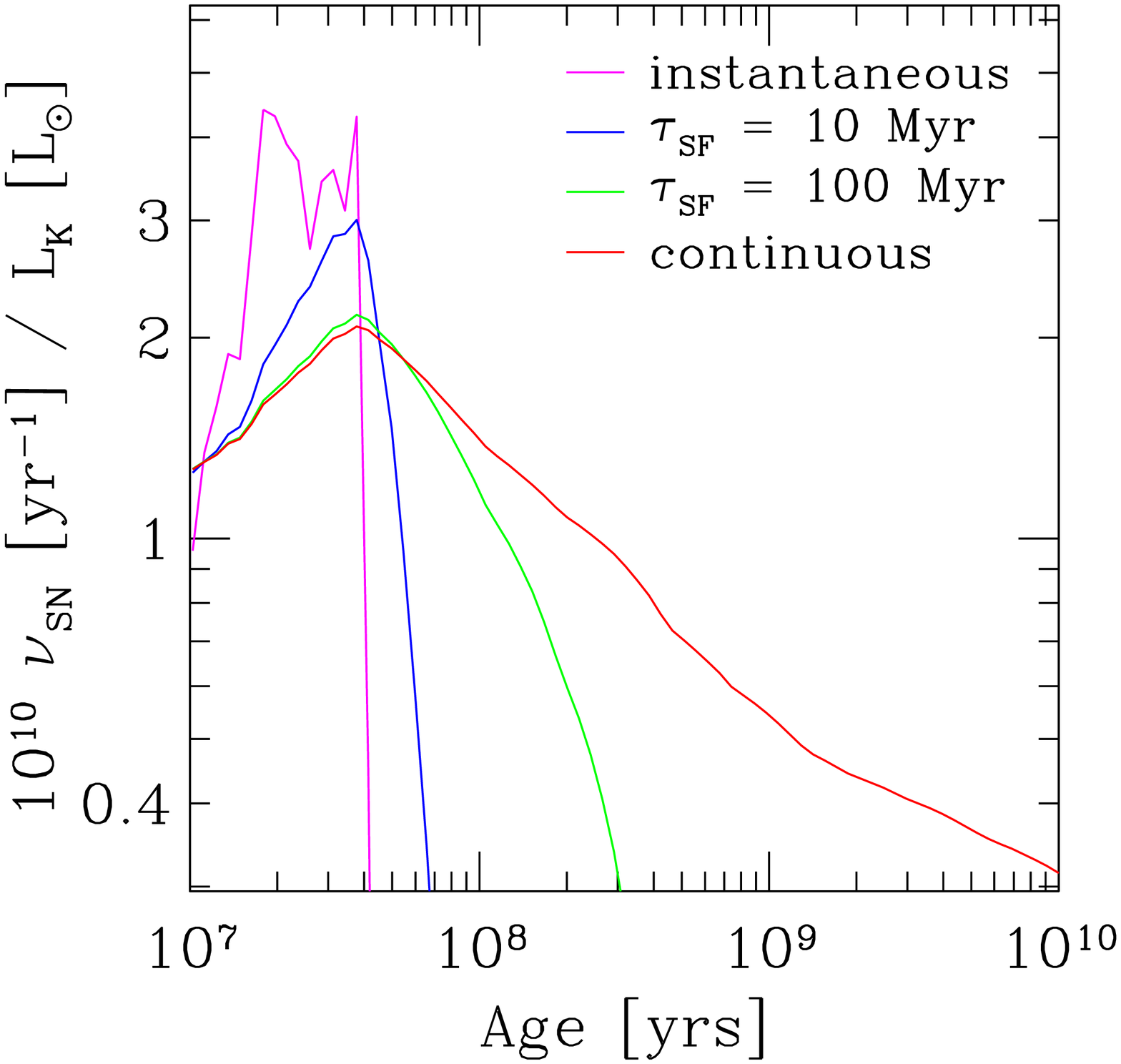} 
\plotone{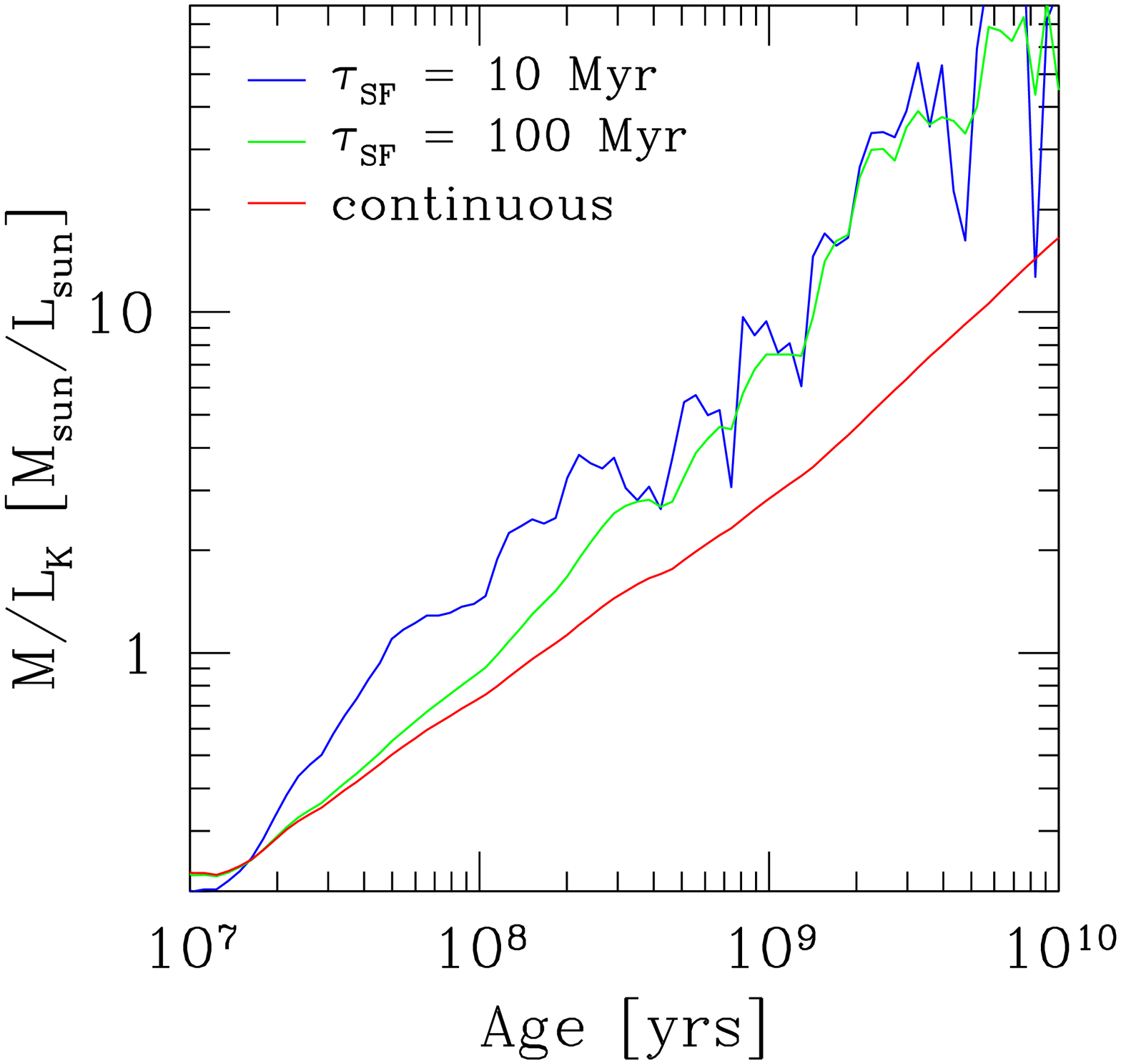} 
\caption{Various diagnostics calcuated with STARS for several star
  formation timescales, as functions of age: Br$\gamma$ equivalent
  width, supernova rate, and mass-to-light ratio. 
Note that all are normalised to the K-band stellar continuum;
and that L$_{\rm K}$ is the total luminosity in the
1.9--2.5\micron\ band in units of bolometric solar luminosity
($1\,L_{\rm bol} = 3.8\times10^{26}\,W$), rather than the other
  frequently used monochromatic definition which has units of the
  solar K-band luminosity density.}
\label{fig:stars}
\epsscale{1.0}
\end{figure}

\begin{figure}
\epsscale{0.7}
\plotone{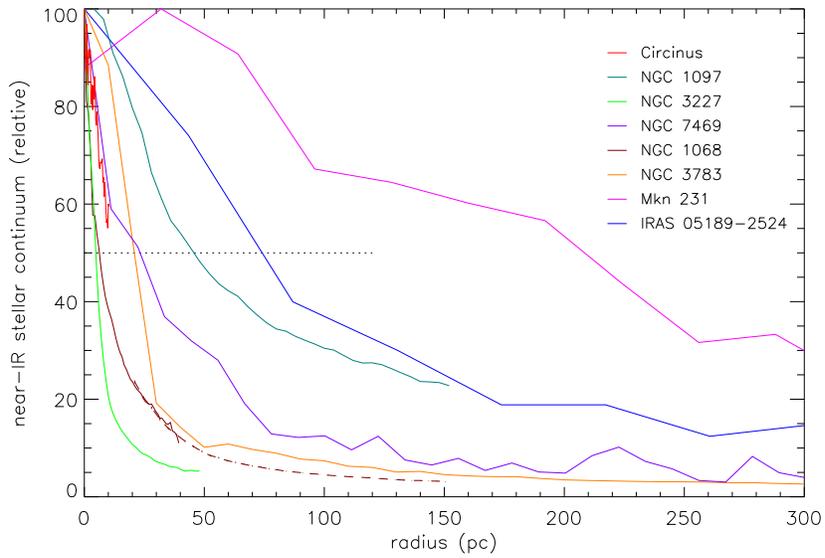} 
\caption{Size scales of nuclear star forming regions. The profile has been
  determined from the CO absorption features in the H or K band, which are
  approximately independent of star formation history (see text for
  details). For longslit data, spatial profiles have been averaged; for
  integral field data, azimuthally averaged proifles are shown. For
  NGC1068, data at two different pixel scales are shown (corresponding
  to the solid and dashed brown lines).
  The horizontal dotted line is drawn at half-maximum height, to
  assist in estimating size scales by eye.
}
\label{fig:sizes}
\epsscale{1.0}
\end{figure}

\begin{figure}
\epsscale{0.87}
\plotone{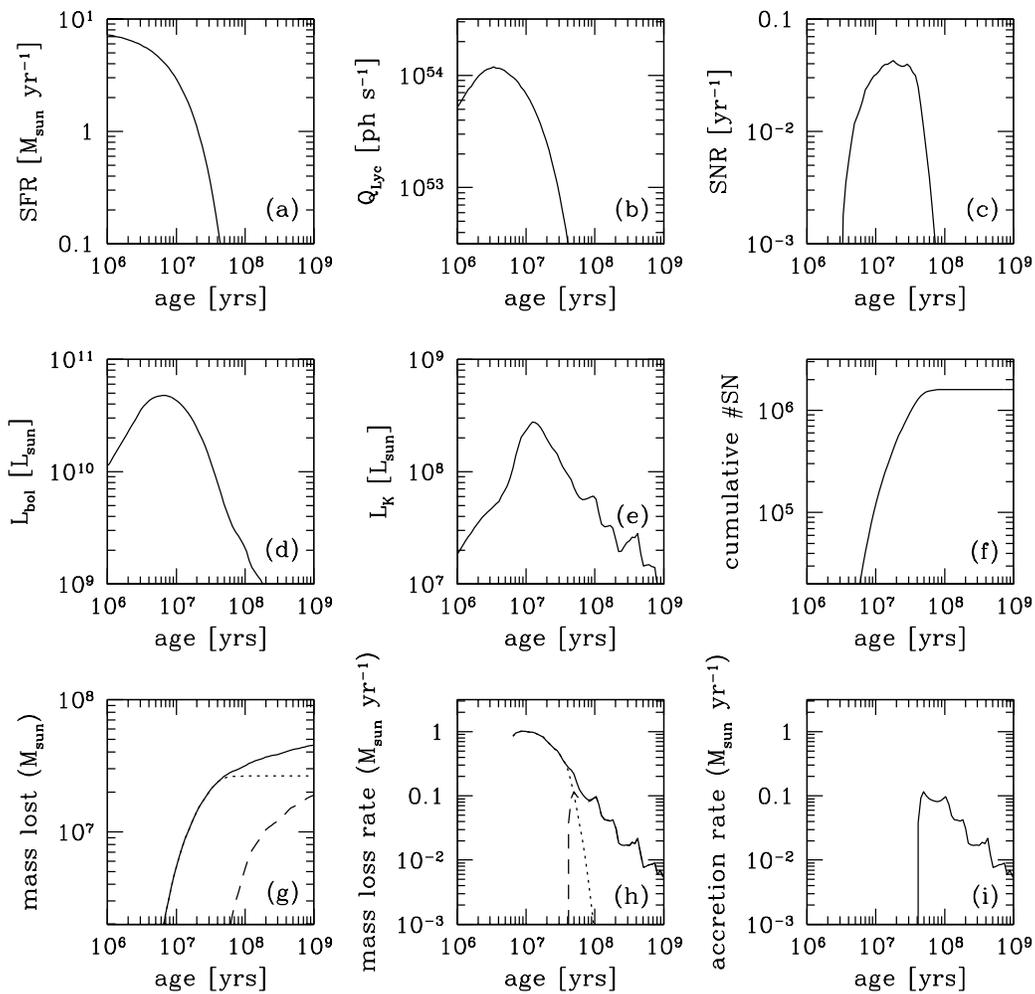} 
\caption{A STARS star formation model based on the main
  characteristics of the observed starbursts, which is illustrative of a
  `typical' nuclear starburst that we have observed.
 The scaling is fixed as
  $2\times10^9$\,M$_\odot$ (typical of that within a 30\,pc radius;
  Fig.~\ref{fig:mag_bol}) at an age of 100\,Myr (the typical age in
  Table~\ref{tab:derprop}). The star formation timescale is 
$\tau_{\rm SF}=10$\,Myr to reproduce a low Br$\gamma$ equivalent width.
The panels are, from top left:
(a) star formation rate (SFR); 
(b) number of ionising photons (Q$_{\rm Lyc}$, proportional to the
  Br$\gamma$ luminosity);
(c) supernova rate (SNR);
(d) bolometric luminosity (L$_{\rm bol}$);
(e) K-band luminosity (L$_{\rm K}$);
(f) cumulative number of supernovae;
(g) cumulative mass that has been recycled back into the ISM by supernovae and
  winds; 
(h) mass loss rate from stars;
(i) rate at which the lost mass can in principle be accreted onto a central
  supermassive black hole (due to its outflow speed; see text for details).
In the last two panels, the mass loss is split into that due
  to OB and Wolf-Rayet stars and supernovae (dotted lines), and that due to
  late-type and AGB stars (dashed lines).
Stellar mass loss in STARS is accounted for at the end of each
     star's life as the difference in mass between the original
     (ZAMS) stellar mass and the remnant mass as the end-product
     of its stellar evolution, as described in \cite{ste98}.
     For this reason, the mass loss rates from OB and Wolf-Rayet
     stars do not appear explicitly in panel (h) at very young
     cluster ages.
}
\label{fig:starstoy}
\epsscale{1.0}
\end{figure}

\begin{figure}
\epsscale{0.65}
\plotone{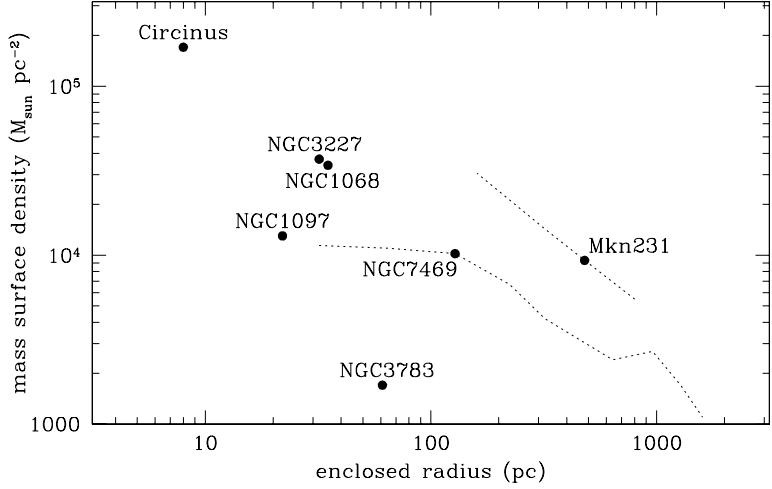} 
\caption{Mean enclosed mass surface density as a function of radius.
The points are from data given in Table~\ref{tab:derprop}; the dashed
lines represent the mass models derived for NGC\,7469 and Mkn\,231
\citep{dav04a,dav04b}.
The galaxies all follow the same trend towards increasing
densities in the central regions.
}
\label{fig:msurfden}
\epsscale{1.0}
\end{figure}

\begin{figure}
\epsscale{0.7}
\plotone{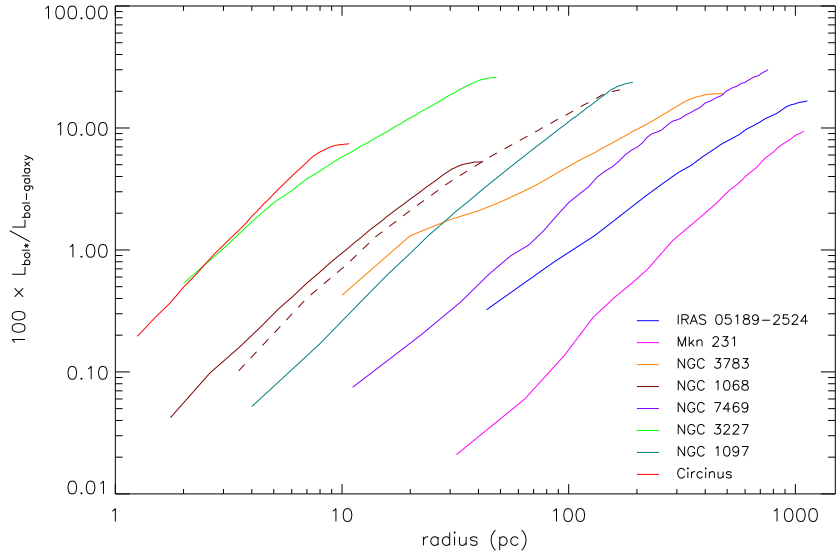} 
\caption{Integrated bolometric luminosity of the young stars 
L$_{\rm bol*}$ as a fraction of that of the galaxy
  L$_{\rm bol-galaxy}$, plotted as a function of radius.
L$_{\rm bol*}$ is calculated from the stellar L$_K$ or L$_H$
assuming that, on the generally small scales here, all the near
infrared stellar continuum originates in the young stars. 
On 10\,pc scales the contribution of young stars is at most a
  few percent of the galaxy's total luminosity, while on kpc scales
  it may be significant and hence comparable to the AGN
  luminosity.
For NGC\,1068, data at two different pixel scales are shown
  (corresponding to the solid and dashed brown lines).
}
\label{fig:mag_agn}
\epsscale{1.0}
\end{figure}

\begin{figure}
\epsscale{0.7}
\plotone{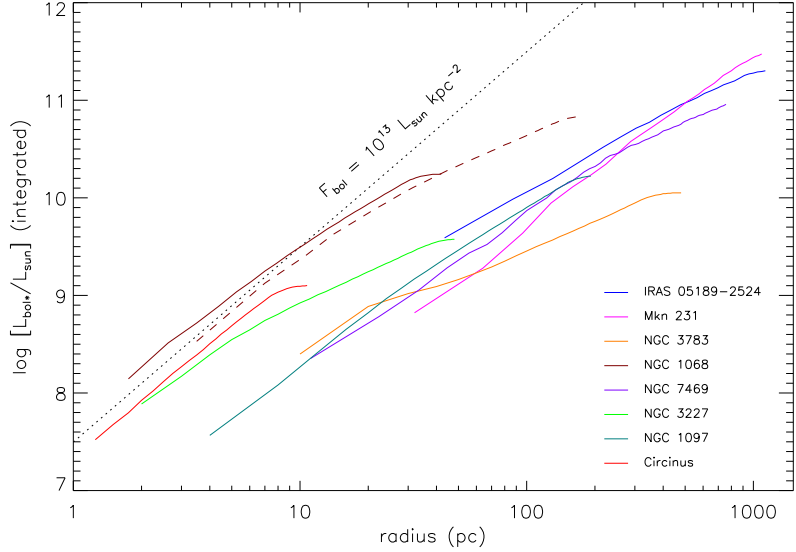} 
\caption{Integrated bolometric luminosity of young stars 
L$_{\rm bol*}$ (see Fig.~\ref{fig:mag_agn}) as a
  function of radius.
For comparison, the dotted line has constant surface brightness.
For NGC\,1068, data at two different pixel scales are shown
  (corresponding to the solid and dashed brown lines).
}
\label{fig:mag_bol}
\epsscale{1.0}
\end{figure}

\begin{figure}
\epsscale{0.4}
\plotone{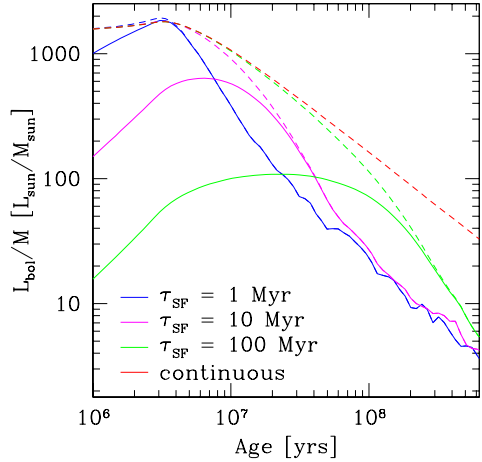} 
\caption{Luminosity to mass ratio calculated by STARS as a function of
  age for different star formation timescales. 
The dotted lines show how the ratio would vary if gas were fed in to a
  cluster at the same rate as it was converted into stars.
The solid lines assume that the gas is present at the
  start, but at the end has all been processed in stars.
A cluster can only exceed 500\,L$_\odot$/M$_\odot$ for a timescale of
  $\sim10$\,Myr.
}
\label{fig:ratio_bol}
\epsscale{1.0}
\end{figure}

\begin{figure}
\epsscale{0.9}
\plotone{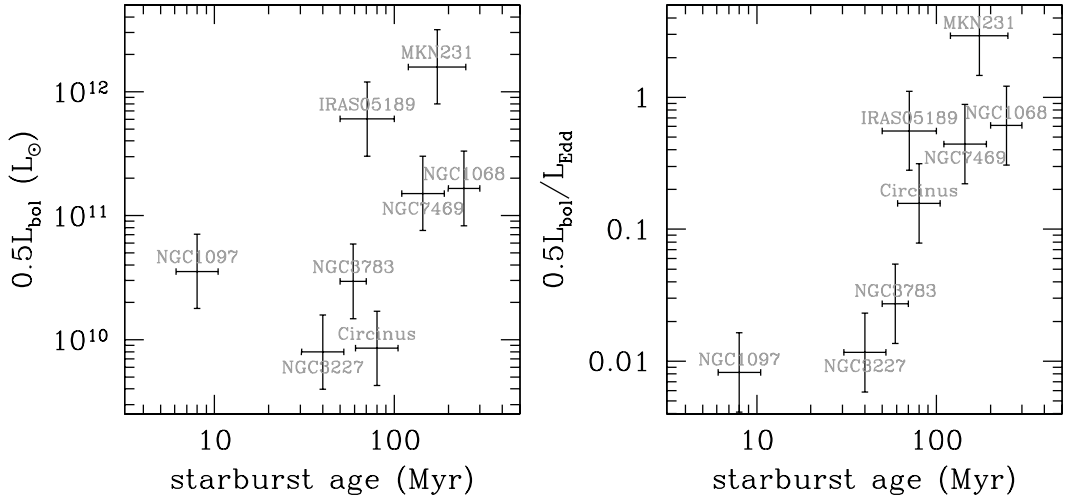} 
\caption{Graph showing how the luminosity of an AGN might be related
  to the age of the most recent episode of nuclear star formation.
On the left is shown the luminosity in solar units; on the right, it
  is with respect to the Eddington luminosity for the black hole.
Generally the luminosity of the AGN is not well known and so we have
  approximated it by 0.5L$_{\rm bol}$, and adopted an uncertainty of
  a factor 2.
The starburst age refers to our best estimate of the most recent episode of
  star formation within the central 10--100\,pc, as given in
  Table~\ref{tab:derprop}. 
See the text for details of the adopted uncertainties.}
\label{fig:age}
\epsscale{1.0}
\end{figure}

\begin{figure}
\epsscale{0.7}
\plotone{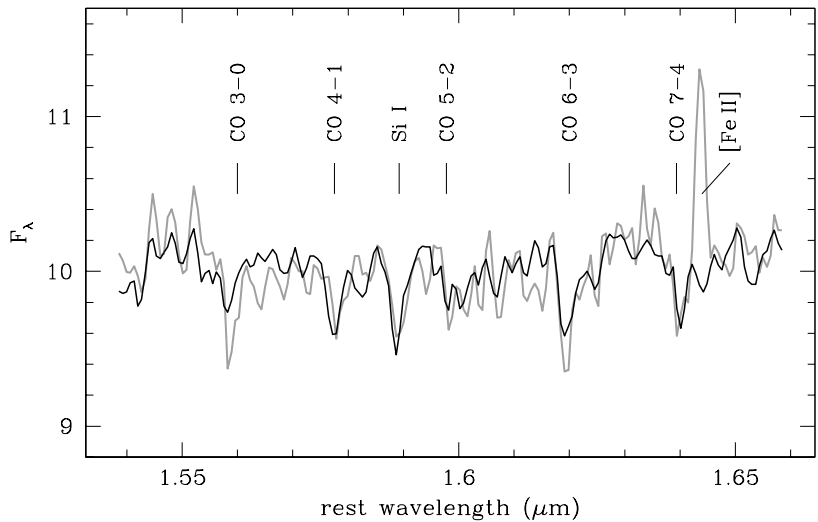} 
\caption{Spectrum (in grey) of IRAS\,05189-2524, integrated over two
  0.22\arcsec\ wide 
  sections centered $\pm$0.27\arcsec\ either side of the nucleus, and which
  have been shifted to match their velocities. Overplotted in black is
  a fit to the 
  continuum, comprising spectra of various supergiant stars. The main
  absorption and emission features have been identified.}
\label{fig:ir05189_spec}
\epsscale{1.0}
\end{figure}

\begin{figure}
\epsscale{1.0}
\plotone{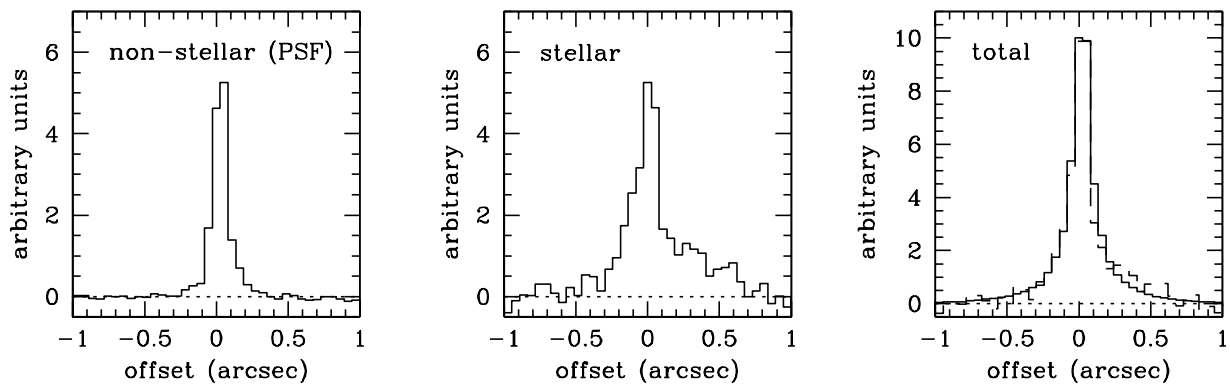} 
\caption{Spatial profiles of non-stellar (left), stellar (centre), and total
  (right) continuum for IRAS\,05189-2524 (1\arcsec\ = 800\,pc).
The first two have been derived at each point along the spatial extent of the
slit from the spectral slope and the stellar
absorption features respectively.
A comparison of their sum (dashed line in right panel) to the total
continuum indicates that the decomposition appears to be reasonable.
}
\label{fig:ir05189_decomp}
\epsscale{1.0}
\end{figure}

\begin{figure}
\epsscale{1.0}
\plotone{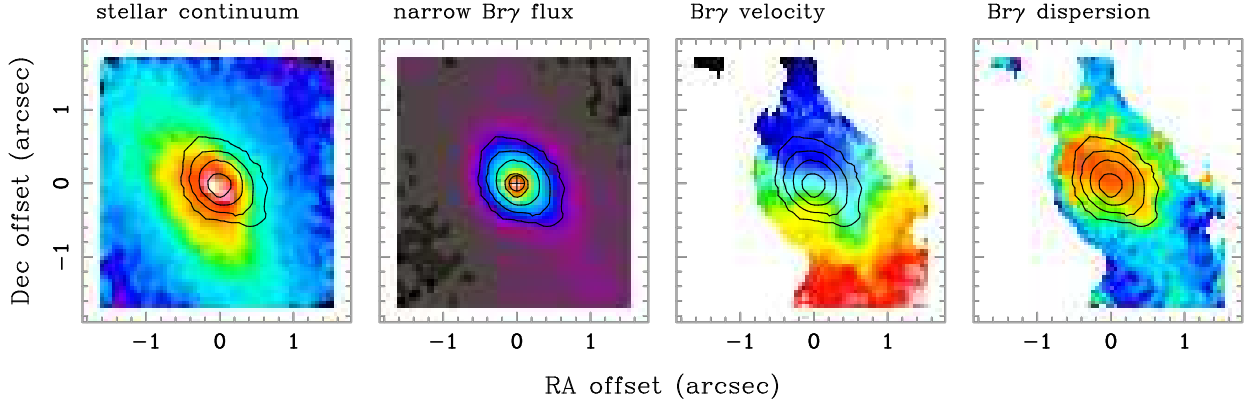} 
\caption{Maps of NGC\,2992 (1\arcsec\ = 160\,pc). 
From left to right: stellar continuum, narrow Br$\gamma$, 
Br$\gamma$ velocity ($-150$ to $+150$\,km\,s$^{-1}$), and
Br$\gamma$ dispersion (0 to 200\,km\,s$^{-1}$).
For reference, on each panel are superimposed contours of the Br$\gamma$ flux
(8, 16, 32, and 64\% of the peak).
The symbol plotted on the map of the line flux indicates the centre of the
broad Br$\gamma$ and non-stellar emission. 
The narrow Br$\gamma$ emission extends far more to the north west than the
stellar continuum.
And, particularly on the south western edge, it exhibits a blue shifted
velocity and high dispersion.
All these are consistent with an interpretation as the apex of an ionisation
cone.}
\label{fig:n2992_brg}
\epsscale{1.0}
\end{figure}

\begin{figure}
\epsscale{0.7}
\plotone{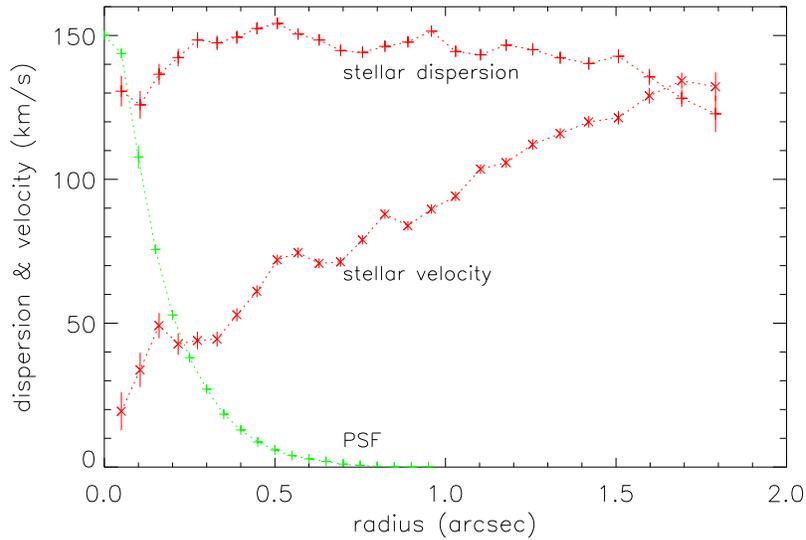} 
\caption{Radial profiles of velocity and dispersion for the stars in
  NGC\,2992 (1\arcsec\ = 160\,pc).
The 2D maps were analysed using the kinemetric technique described by 
  \cite{kra06} which yielded a position angle of $24^\circ$ and an
  inclination of $\sim40^\circ$, not dissimilar to the isophotal
  values of $30^\circ$ and $50^\circ$ respectively.
The rotation curve has been corrected for the inclination.}
\label{fig:n2992_vel}
\epsscale{1.0}
\end{figure}

\begin{figure}
\epsscale{0.7}
\plotone{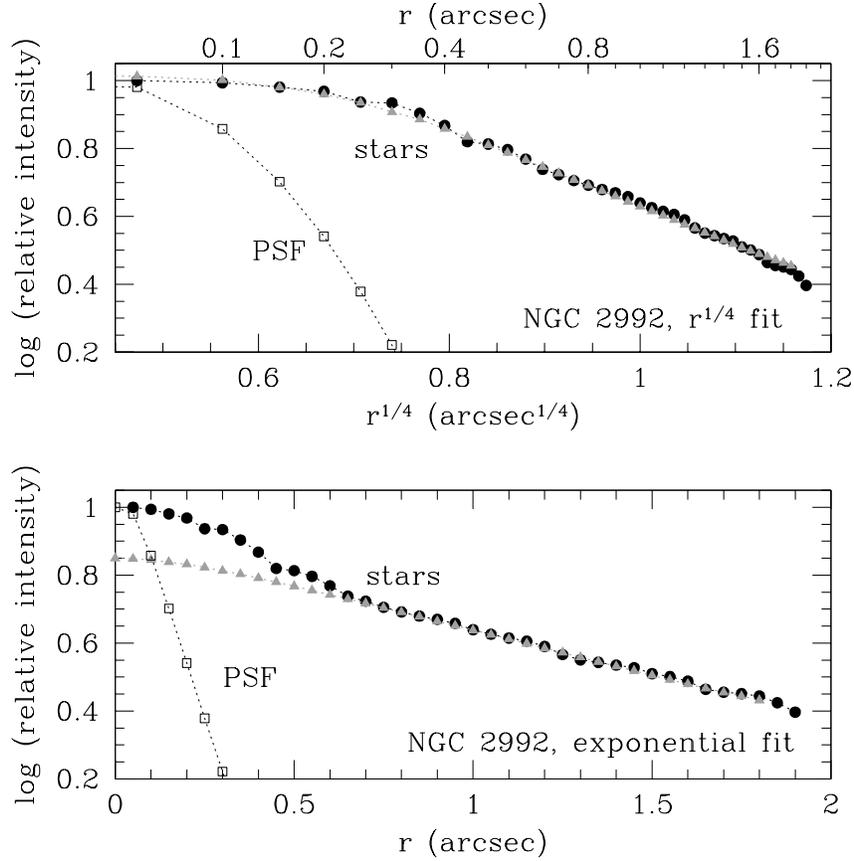} 
\caption{Radial profile of the stellar continuum in NGC\,2992 (1\arcsec\ =
  160\,pc), derived from isophotal analysis.
Solid circles denote the stellar continuum (i.e. already corrected for the
non-stellar component).
Overplotted with triangles are an $r^{1/4}$ law (top panel) and an
  exponential profile (bottom panel).
The profiles were fitted at radii $r>0.5$\arcsec\ and
extrapolated inwards, convolved with the PSF which is shown as open squares.
Both fits are equally good at $r>0.5$\arcsec, but only the exponential
  suggests there might be excess continuum at the centre, arising from
  a distinct stellar population. This is therefore inconclusive.}
\label{fig:n2992_rad}
\epsscale{1.0}
\end{figure}

Taking $V_{\rm rot}=60$\,km\,s$^{-1}$ and $\sigma = 35$\,km\,s$^{-1}$
we derive a dynamical mass of M$_{\rm dyn} = 1.0\times10^8$\,M$_\odot$.
The black hole mass of $3\times10^7$\,M$_\odot$ (from reverberation
mapping, \citealt{pet04}) is only 30\% of this,
and so cannot be dominating the dynamics on this scale unless its mass
is underestimated.
With respect to this, we note that \cite{pet04} claim the statistical
uncertainty in 
masses derived from reverberation mapping is about a factor 3.
Alternatively, there may be a compact mass of gas and stars at
$r<0.3$\arcsec.
However, including $\sigma$ in the mass estimate implicitly assumes
that the dispersion arises from macroscopic motions.
On the other hand, because we are observing
only the hot H$_2$, it is possible that the dispersion is dominated by
turbulence arising from shocks or UV heating of clouds that generate
the 1-0\,S(1) emission -- issues that are discussed in more detail by
Hicks et al. (in prep.).
In this case we will have overestimated the dynamical mass.
Excluding $\sigma$ from the mass estimation yields 
$M_{\rm dyn}=5\times10^7$\,M$_\odot$.
We consider these two estimates as denoting the maximum range of
possible masses.
Subtracting M$_{\rm BH}$ then gives a mass of stars and gas in the
range (2--7)$\times10^7$\,M$_\odot$, implying a mass surface density of
1700--6000\,M$_\odot$\,pc$^{-2}$ and 
$M/L_K = 0.6$--2.1\,M$_\odot$/L$_\odot$.
Based on these ratios alone, Fig~\ref{fig:stars} indicates that the
characteristic age of the star formation may be as low as
$\sim70$\,Myr, although it could also be an order of magnitude greater.
Without additional diagnostics we cannot discriminate further.

We are unable to use Br$\gamma$ as an additional constraint on the star
formation history.
Its morphology and velocity field are similar to that of
[Si{\sc vi}], and rather different from the 1-0\,S(1).
It shows an extension to the north which appears to be outflowing at
$>50$\,km\,s$^{-1}$ (Fig~\ref{fig:n3783_maps}) -- perhaps tracing
an ionisation cone.
Since the Br$\gamma$ resembles the [Si{\sc vi}], it is reasonable to
conclude that it too is associated with the AGN rather than star
formation.
Thus the equivalent width of Br$\gamma$ (with respect to the stellar
continuum) of $W_{\rm Br\gamma}=30$\AA\ represents an upper limit to
that associated with star formation.

The radio continuum in the nucleus of NGC\,3783 has been measured with
several beam sizes at 8.5\,GHz.
For a beam of $1.59\arcsec\times0.74\arcsec$, \cite{mor99}
found it was unresolved with a flux density of $8.15\pm0.24$\,mJy.
With a smaller $\sim0.25$\arcsec\ beam, \cite{sch01} measured a total
flux density of 8.0\,mJy dominated by an unresolved component of
$7.7\pm0.05$\,mJy.
At smaller scales still of $\sim0.03$\arcsec\ corresponding to 6\,pc,
\cite{sad95} placed an upper limit on the 8.5\,GHz flux density of
7\,mJy.
Taken together, these results imply that there is some modest 8.5\,GHz
radio continuum of 0.7--1\,mJy extended on scales of 0.3--1\arcsec.
Based on this we estimate a supernova rate as described in
Section~\ref{sec:diag} of $\sim0.007$\,yr$^{-1}$, and hence a 
ratio $10^{10}\,\nu_{\rm SN}/L_K \sim 2$.
Given that the unresolved radio continuum on the smallest scales is an
upper limit, the extended component may be stronger and hence the true
$\nu_{\rm SN}/L_K$ ratio may be greater than that estimated here.
Fig~\ref{fig:stars} then puts a relatively strong limit of
$\sim50$\,Myr on the maximum age of the star formation.

This age is fully consistent with that above associated with our lower
mass estimate.
The value of $W_{\rm Br\gamma}<30$\AA\ above does not impose additional
constraints, although we note that 
if the Br$\gamma$ flux associated with star formation is only a small
fraction of the total then it would imply that the
timescale over which the star formation was active is no longer than
a few times $\sim10$\,Myr.
Therefore in the nucleus ($r<0.3$\arcsec) of NGC\,3783 we adopt
50--70\,Myr as the age 
of the star formation and $M_{\rm dyn}=2\times10^7$\,M$_\odot$ as the
dynamical mass excluding the central supermassive black hole.



\clearpage

\begin{deluxetable}{llccclrl}
\tablecaption{Table of Observations\label{tab:obs}} 
\tablewidth{0pt}
\tablehead{
\colhead{Object} &
\colhead{Band\tablenotemark{a}} & 
\colhead{Res.\tablenotemark{b} (\arcsec)} & 
\colhead{Date} & 
\colhead{Instrument}
}

\startdata

Mkn\,231\tablenotemark{c}  & H & 0.176 & May '02 & Keck, NIRC2 \\
NGC\,7469\tablenotemark{c} & K & 0.085 & Nov '02 & Keck, NIRSPAO \\
                           & K & 0.15  & Jul '04 & VLT, SINFONI \\
Circinus\tablenotemark{c}  & K & 0.22  & Jul '04 & VLT, SINFONI \\
NGC\,3227\tablenotemark{c} & K & 0.085 & Dec '04 & VLT, SINFONI \\
IRAS\,05189-2524           & H & 0.12  & Dec '02 & VLT, NACO \\
NGC\,2992                  & K & 0.30  & Mar '05 & VLT, SINFONI \\
NGC\,1097                  & H & 0.245 & Oct '05 & VLT, SINFONI \\
NGC\,1068                  & H & 0.10  & Oct '05 & VLT, SINFONI \\
                           & H & 0.13  & Dec '02 & VLT, NACO \\
NGC\,3783                  & K & 0.17  & Mar '05 & VLT, SINFONI \\

\enddata

\tablenotetext{a}{Band used for determining the quantitative star
  formation properties. NGC\,1097, NGC\,1068, and NGC\,3783 were
  actually observed in H and K bands.}

\tablenotetext{b}{Spatial resolution (FWHM) estimated from the data
  itself, using the methods described in Section~\ref{sec:obs}.}

\tablenotetext{c}{References to detailed studies of individual objects: 
Mkn 231 \citep{dav04b}, 
NGC 7469 \citep{dav04a}, 
Circinus, \citep{mul06},
NGC 3227 \citep{dav06}.}

\end{deluxetable}


\begin{deluxetable}{llcccccc}
\tablecaption{Summary of basic data for AGN\label{tab:basicdata}} 
\tablewidth{0pt}
\tablehead{
\colhead{Object} &
\colhead{Classification\tablenotemark{a}} &
\colhead{Distance} &
\colhead{log$\frac{L_{\rm bol}}{L_\odot}$\tablenotemark{b}} & 
\colhead{log$\frac{M_{\rm BH}}{M_\odot}$} &
\colhead{Ref.\tablenotemark{c}} \\
\colhead{} &
\colhead{} &
\colhead{Mpc} &
\colhead{} &
\colhead{} &
\colhead{for M$_{\rm BH}$}
}

\startdata

Mkn\,231         & ULIRG, Sy\,1, QSO &       170 & 12.5 & 7.2 & 1 & \\
NGC\,7469        & Sy\,1             &    \phn66 & 11.5 & 7.0 & 2 & \\
Circinus         & Sy\,2             & \phn\phn4 & 10.2 & 6.2 & 3 & \\
NGC\,3227        & Sy\,1             &    \phn17 & 10.2 & 7.3 & 4 & \\
IRAS\,05189-2524 & ULIRG, Sy\,1      &       170 & 12.1 & 7.5 & 1 & \\
NGC\,2992        & Sy\,1             &    \phn33 & 10.7 & 7.7 & 5 & \\
NGC\,1097        & LINER, Sy\,1      &    \phn18 & 10.9 & 8.1 & 6 & \\
NGC\,1068        & Sy\,2             &    \phn14 & 11.5 & 6.9 & 7 & \\
NGC\,3783        & Sy\,1             &    \phn42 & 10.8 & 7.5 & 2 & \\

\enddata

\tablenotetext{a}{\,Classifications are taken primarily from the NASA/IPAC
  Extragalactic Database. In addition, we have labelled as Seyfert~1 those for
  which we have observed broad (i.e. FWHM $>$ 1000\,km\,s$^{-1}$)
  Br$\gamma$; see also Fig~\ref{fig:bbrg}.}

\tablenotetext{b}{\,Calculated in the range 8--1000\,$\mu$m from the
  IRAS 12--100\,$\mu$m flux densities; with an additional correction
  for optical and near-infrared luminosity in cases where appropriate.} 

\tablenotetext{c}{\,References for black hole masses: 
(1) \cite{das06b};
(2) \cite{pet04};
(3) \cite{gre03};
(4) \cite{dav06};
(5) \cite{woo02};
(6) \cite{lew06}; 
(7) \cite{lod03}
}

\end{deluxetable}


\clearpage
\thispagestyle{empty}
\begin{deluxetable}{lcccccccccc}
\rotate
\tablecaption{Measured \& Derived Properties of the Nuclei\tablenotemark{a}\label{tab:derprop}} 
\tablewidth{0pt}
\tablehead{

\colhead{Object} &
\multicolumn{2}{c}{radius} &
\colhead{log$\frac{L_K*}{L_\odot}$} & 
\colhead{log$\frac{M_{\rm dyn}}{M_\odot}$} &
\colhead{$\Sigma_{\rm dyn}$} &
\colhead{W$_{\rm Br\gamma}$} & 
\colhead{$M_{\rm dyn}/L_K$} &
\colhead{$10^{10} \nu_{SN}/L_K$} &
\colhead{age} &
\colhead{$\langle$SFR$\rangle$}
\\
\colhead{} &        
\colhead{\arcsec} & 
\colhead{pc} &      
\colhead{} &        
\colhead{} &        
\colhead{$10^4\,M_\odot\,pc^{-2}$} & 
\colhead{\AA} &     
\colhead{$M_\odot/L_\odot$} & 
\colhead{yr$^{-1}$\,$L_\odot^{-1}$} & 
\colhead{Myr} &     
\colhead{$M_\odot$\,yr$^{-1}$\,kpc$^{-2}$} 
}

\startdata

Mkn\,231\tablenotemark{b}  & 0.6\phn &       480 & 9.3 & 9.8 & \phn0.9    &   \phm{$<$}--- & \phn3.1    & \phm{$<$}20\phd\phn    & 120--250 & 25--50    \\
NGC\,7469                  & 0.4\phn &       128 & 8.5 & 8.7 & \phn1.0    &    \phm{$<$}11 & \phn1.6    & \phm{$<$}\phn3\phd\phn & 110--190 & 50--100   \\
Circinus                   & 0.4\phn & \phn\phn8 & 6.2 & 7.5 & 17\phd\phn &    \phm{$<$}30 & 23\phd\phn & \phm{$<$}\phn1.5       &  80      & $\sim70$  \\
NGC\,3227\tablenotemark{c} & 0.4\phn &    \phn32 & 7.8 & 8.0 & \phn3.7    & \phn\phm{$<$}4 & \phn1.9    & \phm{$<$}\phn2.2       &  40      & $\sim380$ \\
IRAS\,05189-2524           & 0.55    &       450 & 9.3 & --- & ---        & \phn\phm{$<$}4 & ---        & \phm{$<$}\phn5\phd\phn &  50--100 & 30--70    \\
NGC\,2992\tablenotemark{d} & 0.4\phn &    \phn64 & 7.5 & --- & ---        &          $<$12 & ---        & \phm{$<$}\phn1\phd\phn & ---      & ---       \\
NGC\,1097\tablenotemark{e} & 0.25    &    \phn22 & 6.7 & 8.2 & \phn1.3    & \phn\phm{$<$}1 & \phn4.5    & \phm{$<$}\phn1.4       &  8       & $\sim80$  \\
NGC\,1068                  & 0.5\phn &    \phn35 & 7.6 & 8.1 & \phn3.4    & \phn\phm{$<$}4 & \phn3.0    & $<$20\phd\phn          & 200--300 & 90--170   \\
NGC\,3783\tablenotemark{f} & 0.3\phn &    \phn60 & 7.5 & 7.3 & \phn0.2    &          $<$30 & \phn0.6    & \phm{$<$}\phn2\phd\phn &  50--70  & 30--60    \\

\enddata

\tablenotetext{a}{\,The methods used to measure these quantities
  (within the radii given) are described
in Section~\ref{sec:diag}. Specific issues associated with individual objects
are discussed in Appendix~\ref{sec:obj}.}

\tablenotetext{b}{\,M$_{\rm dyn}$ depends strongly on even small changes to the inclination; 
  here it is given for $i=10^\circ$. 
  Correcting M$_{\rm dyn}$ for an estimate of the gas mass given
  in \cite{dow98} yields $M/L_K = 2.3 M_\odot/L_\odot$.}

\tablenotetext{c}{\,The best star formation models indicate that $M/L_K$ is much less than the limit given here using the dynamical mass.}

\tablenotetext{d}{\,It is likely that much of the narrow Br$\gamma$ in
  the nuclear region here is associated with an ionisation cone.
In addition, the high stellar velocity dispersion, even on the
  smallest scales we have been able to measure, suggests
  that the K-band light is dominated by the bulge.}

\tablenotetext{e}{\,Correcting $W_{\rm Br\gamma}$ for the old stellar
  population would probably yield a value in the range 2--5\AA.
Even on this scale the dynamical mass is dominated by the supermassive
  black hole. Both $\Sigma_{\rm dyn}$ and $M/L_K$ are estimated after 
subtracting $M_{\rm BH}$.}

\tablenotetext{f}{\,Much of the Br$\gamma$ here is outflowing and hence associated with the AGN.
M$_{\rm dyn}$ is derived from gas kinematics as described in the text.
Both $\Sigma_{\rm dyn}$ and $M/L_K$ are estimated after subtracting $M_{\rm BH}$.}

\end{deluxetable}



\clearpage

\clearpage

\begin{figure}
\epsscale{0.7}
\plotone{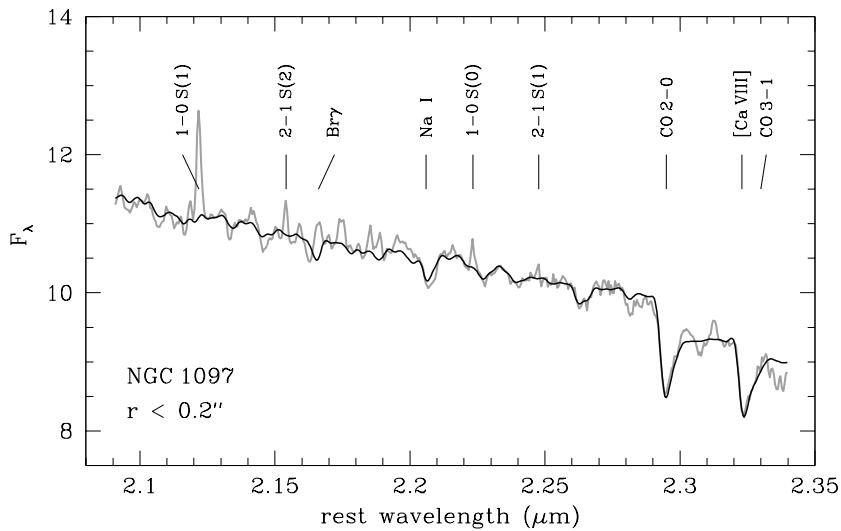}
\caption{Spectrum of NGC\,1097 (thick grey line), extracted within an
  aperture of radius 0.2\arcsec\ and scaled arbitrarily.
Overdrawn (thin black line) is a match to the stellar continuum
  constructed from template spectra of several late-type supergiant
  stars and a blackbody function representing the non-stellar component. 
Notably, Br$\gamma$ in the nucleus is extremely weak even in the nucleus.}
\label{fig:n1097_spec}
\epsscale{1.0}
\end{figure}


\begin{figure}
\epsscale{0.7}
\plotone{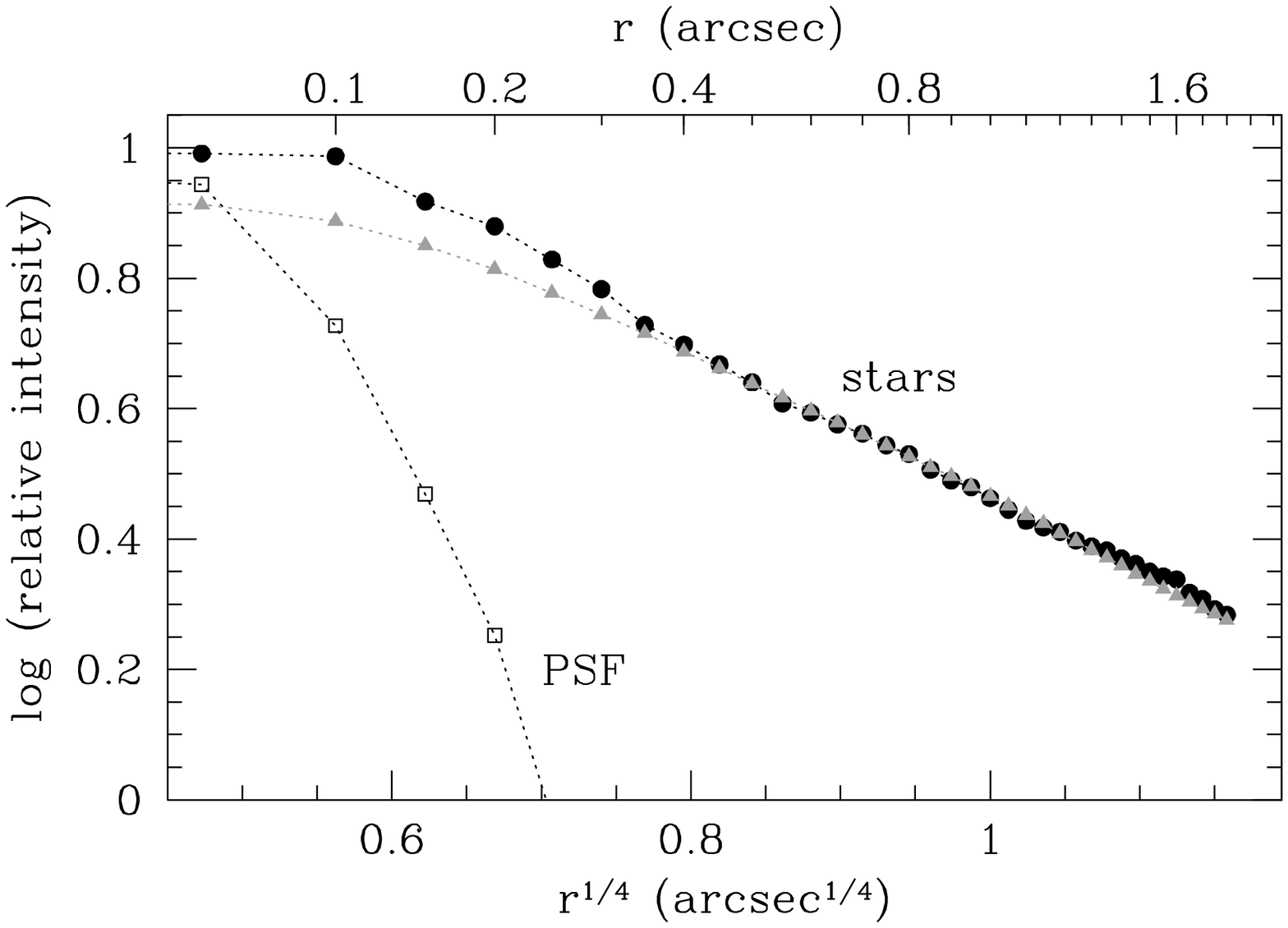}
\caption{Radial profile of the stellar continuum in NGC\,1097 (1\arcsec\ =
  80\,pc).
The solid circles denote the stellar continuum (i.e. already corrected for the
non-stellar component).
The triangles denote an $r^{1/4}$ profile fitted to radii $r>0.5$\arcsec\ and
extrapolated inwards.
This model has been convolved with the PSF, shown as open squares for
comparison.
Note that even though an exponential profile might match the data
  equally well, an $r^{1/4}$ profile provides a stronger
  constraint on whether there is excess continuum at the centre.
}
\label{fig:n1097_prof}
\epsscale{1.0}
\end{figure}


\begin{figure}
\epsscale{0.65}
\plotone{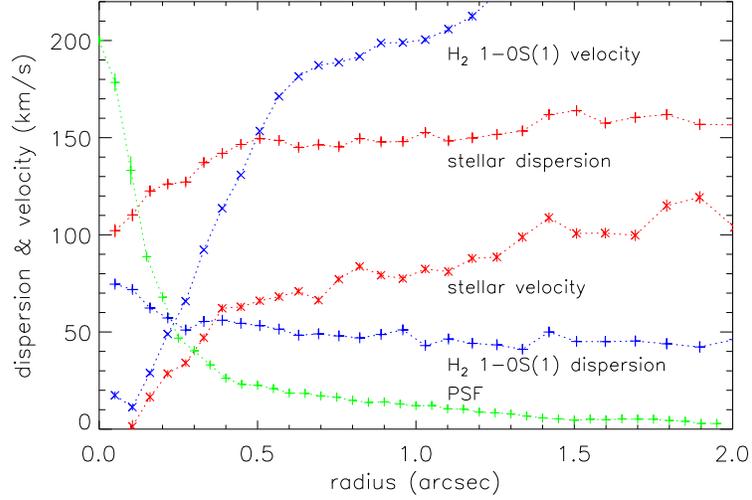} 
\caption{Radial profiles of velocity and dispersion for the gas and stars in
  NGC\,1097 (1\arcsec\ = 80\,pc).
The 2D maps were created by convolving template spectra (i.e unresolved line
profile for the gas, stellar template for the stars) with a Gaussian and
minimising the difference with respect to the galaxy spectrum at each spatial
pixel. 
These were then analysed using the kinemetric technique described by
  \cite{kra06} which yielded the same position angle of $-49^\circ$
  for the gas and stars, and similar inclinations of $32^\circ$ and
  $43^\circ$ respectively. 
The rotation curve has been corrected for the inclination.}
\label{fig:n1097_dispvel}
\epsscale{1.0}
\end{figure}


\begin{figure}
\epsscale{0.8}
\plotone{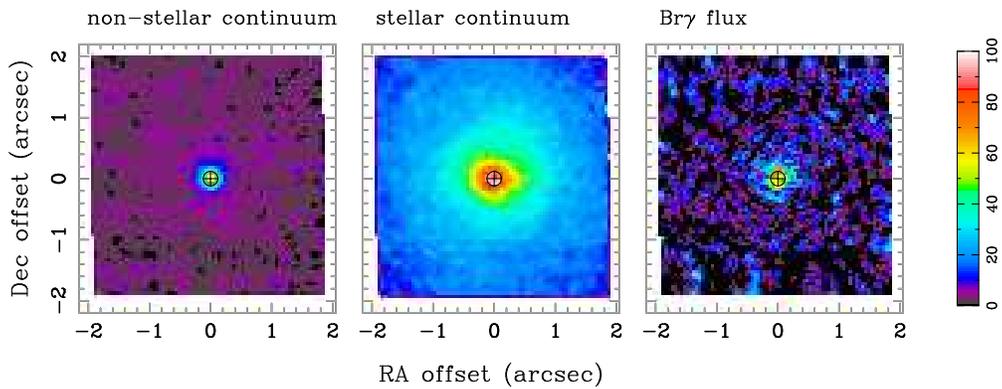} 
\caption{Maps of K-band non-stellar continuum (left), stellar continuum
  (centre), and Br$\gamma$ line flux (right) for the central few arcsec of
  NGC\,1097 (1\arcsec\ = 80\,pc).
In each case, the centre (as defined by the non-stellar continuum) is marked
by a crossed circle.
The colour scale is shown on the right, as percentage of the peak in each
  map.}
\label{fig:n1097_maps}
\epsscale{1.0}
\end{figure}


\begin{figure}
\epsscale{0.65}
\plotone{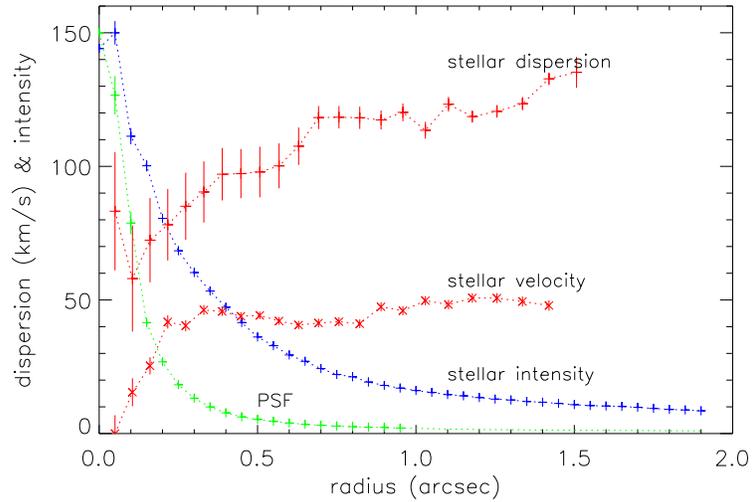} 
\caption{Radial profiles of velocity and dispersion for stars in
  NGC\,1068 (1\arcsec\ = 70\,pc).
The 2D maps were then analysed using kinemetry \cite{kra06}, yielding an
  inclination of $40^\circ$ and a position angle of $85^\circ$.
The rotation curve has been corrected for the inclination.
Also plotted for comparison are azimuthally averaged radial profiles
of the H-band stellar luminosity and the PSF.}
\label{fig:n1068_prof}
\epsscale{1.0}
\end{figure}


\begin{figure}
\epsscale{0.65}
\plotone{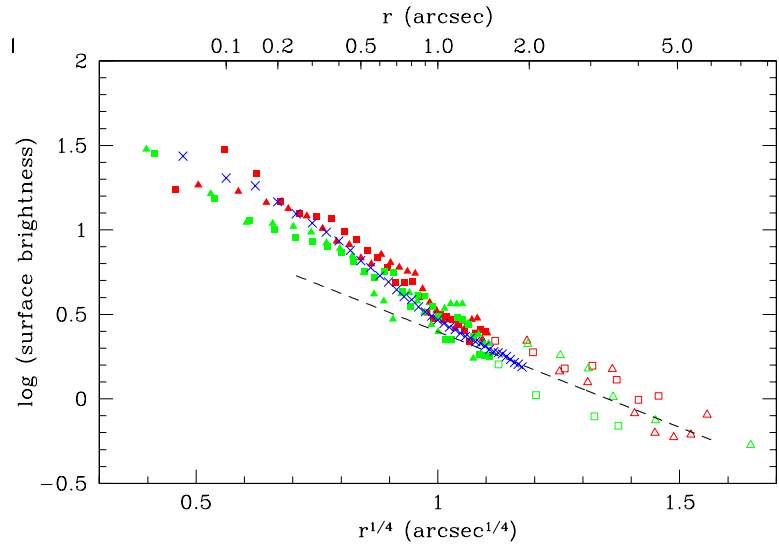} 
\caption{Radial profiles of the H-band stellar luminosity in
  NGC\,1068 (1\arcsec = 70\,pc) from NACO longslit data at position
  angle $0^\circ$ and $90^\circ$ (green and red symbols; squares and
  triangles deonte opposite sides of the nucleus).
To trace the profile to larger radius, open symbols are each the mean of
9 points.
Superimposed are SINFONI data from Fig.~\ref{fig:n1068_prof} (blue
crosses, flux scaled to match). 
The dashed line denotes an $r^{1/4}$ law with 
$R_{\rm eff}=1.5$\arcsec\ to match the outer profile. 
At $r<1$\arcsec, the stellar continuum reveals an excess above the
  inward extrapolation of this profile.}
\label{fig:n1068_prof_wide}
\epsscale{1.0}
\end{figure}


\begin{figure}
\epsscale{1.0}
\plotone{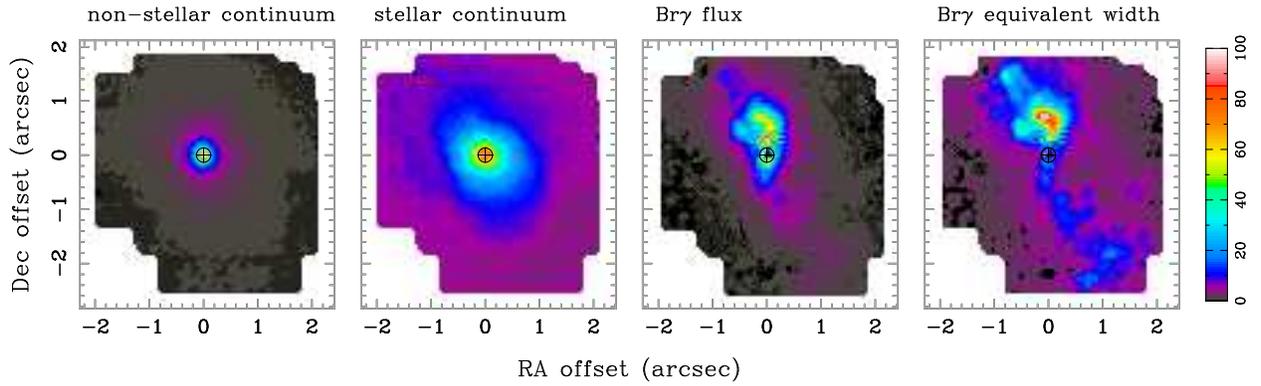} 
\caption{Maps of the central few arcsec of NGC\,1068 (1\arcsec\ = 70\,pc):
H-band non-stellar continuum (far left) and stellar continuum (centre left);
also Br$\gamma$ line flux (center right) and Br$\gamma$ equivalent width (far
right).
In each case, the centre (as defined by the non-stellar continuum) is marked
by a crossed circle.
The colour scale is shown on the right, as percentage of the peak in each
  map (and also as $W_{\rm Br\gamma}$ in \AA).}
\label{fig:n1068_maps}
\epsscale{1.0}
\end{figure}


\clearpage

\begin{figure}
\epsscale{0.8}
\plotone{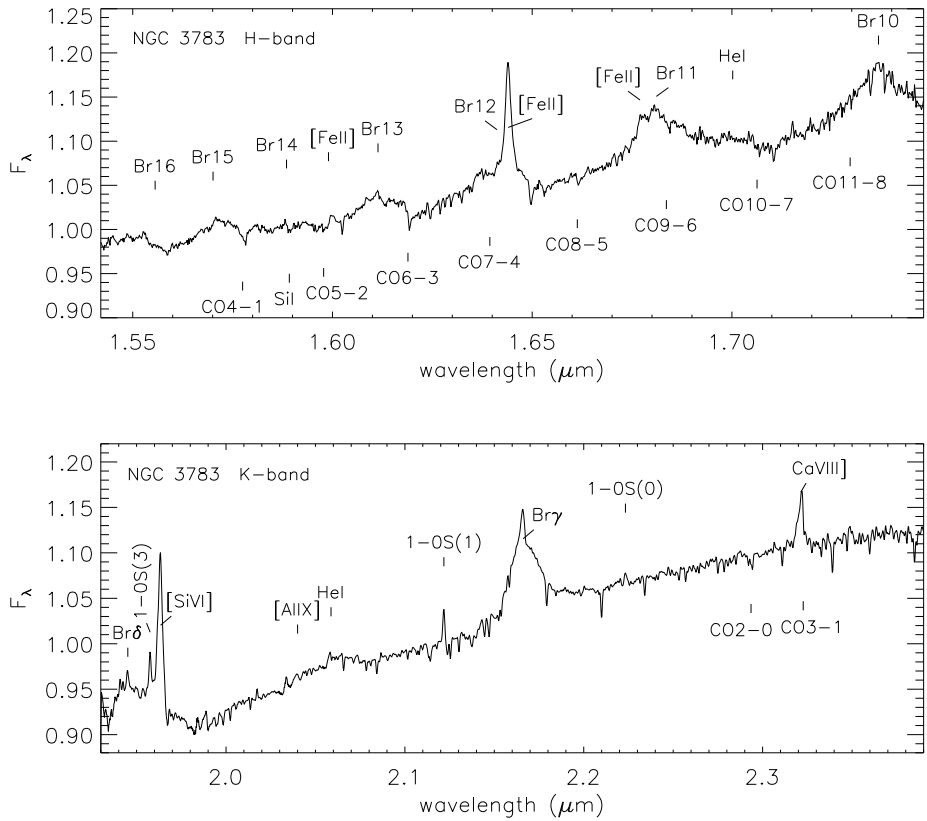} 
\caption{H- and K-band spectra of the central 1\arcsec\ of NGC\,3783,
  with the prominent emission and absorption features labelled.
In the H-band it is challenging to measure the stellar absorption due
  to the very strong brackett emission from the AGN's broad line
  region.
Instead we have used the K-band CO\,2-0 bandhead even though the dilution
  at this wavelength is extreme.
}
\label{fig:n3783_spec}
\epsscale{1.0}
\end{figure}


\begin{figure}
\epsscale{0.65}
\plotone{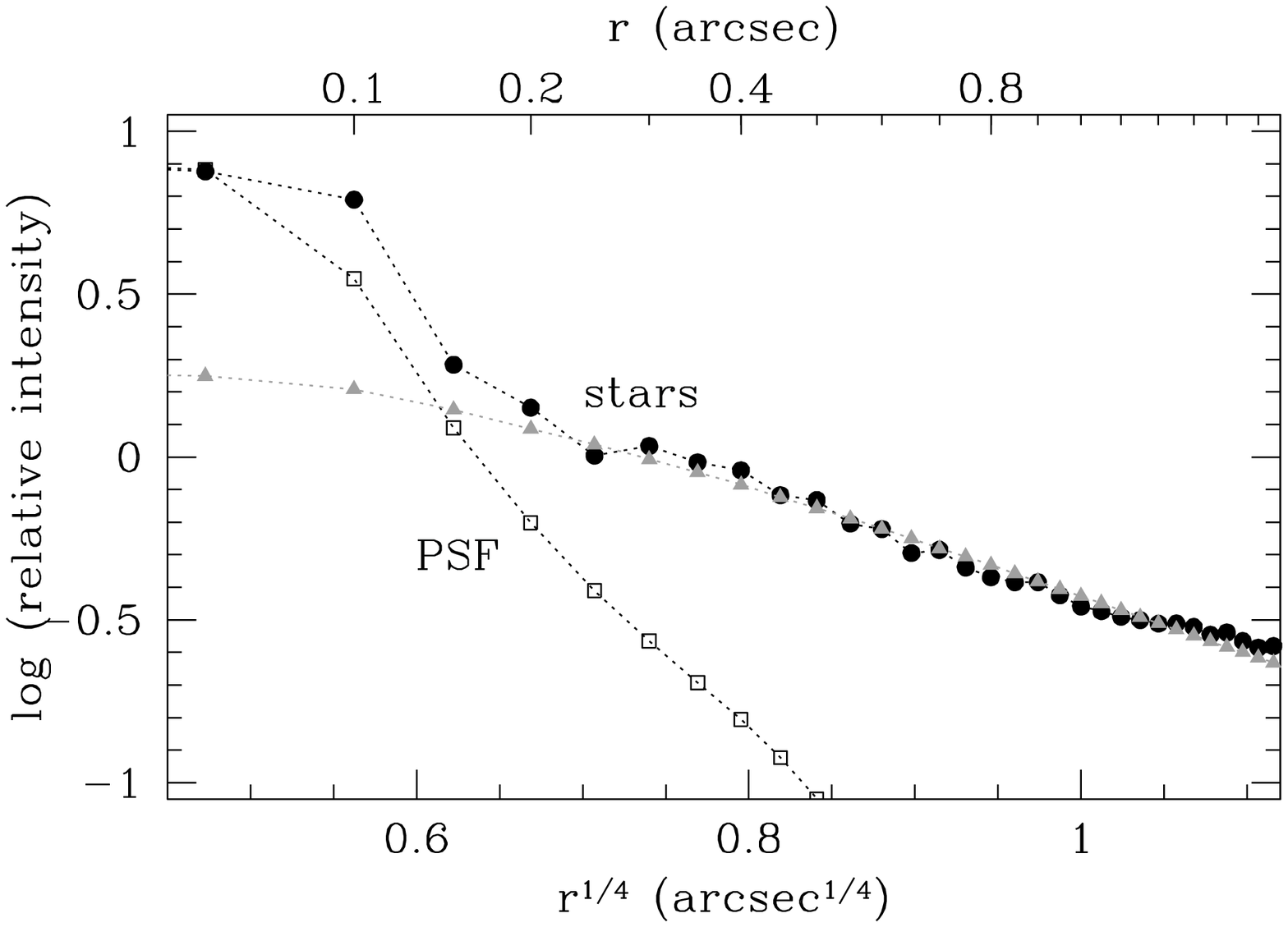} 
\caption{Radial profile of the stellar continuum in NGC\,3783 
(1\arcsec\ = 200\,pc).
The solid circles denote the stellar continuum (i.e. already corrected for the
non-stellar component).
The triangles denote an $r^{1/4}$ profile fitted to radii
$0.2<r<1.6$\arcsec\ and extrapolated inwards.
This model has been convolved with the PSF, shown as open squares for
comparison.
Note that even though an exponential profile might match the data
  equally well, an $r^{1/4}$ profile provides a stronger
  constraint on whether there is excess continuum at the centre.
}
\label{fig:n3783_rad}
\epsscale{1.0}
\end{figure}


\begin{figure}
\epsscale{0.65}
\plotone{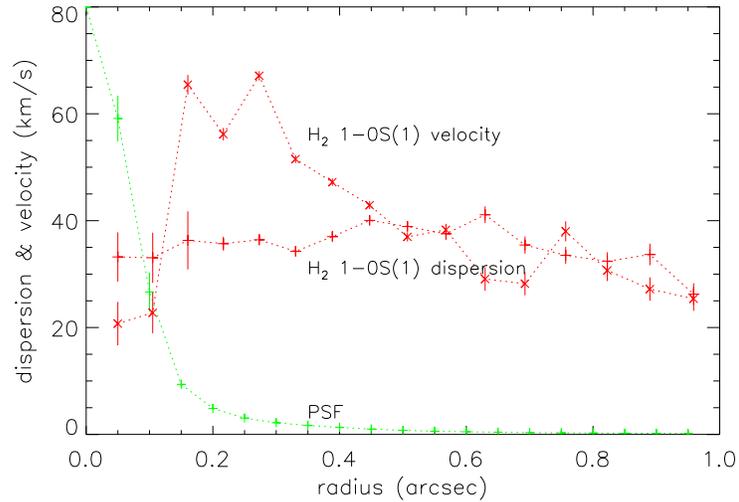} 
\caption{Rotation curve derived from the the H$_2$ 1-0\,S(1) velocity
  field in NGC\,3783 (1\arcsec\ = 200\,pc).
Also shown is the dispersion as a function of radius.
The velocity field was analysed using kinemetry \citep{kra06} which
  yielded a major axis of about $-14^\circ$ and an inclination in the
  range 35--39$^\circ$.
The drop in velocity at $r<0.15$\arcsec\ maybe due to beam smearing
  across the nucleus.
}
\label{fig:n3783_rot}
\epsscale{1.0}
\end{figure}


\begin{figure}
\epsscale{0.7}
\plotone{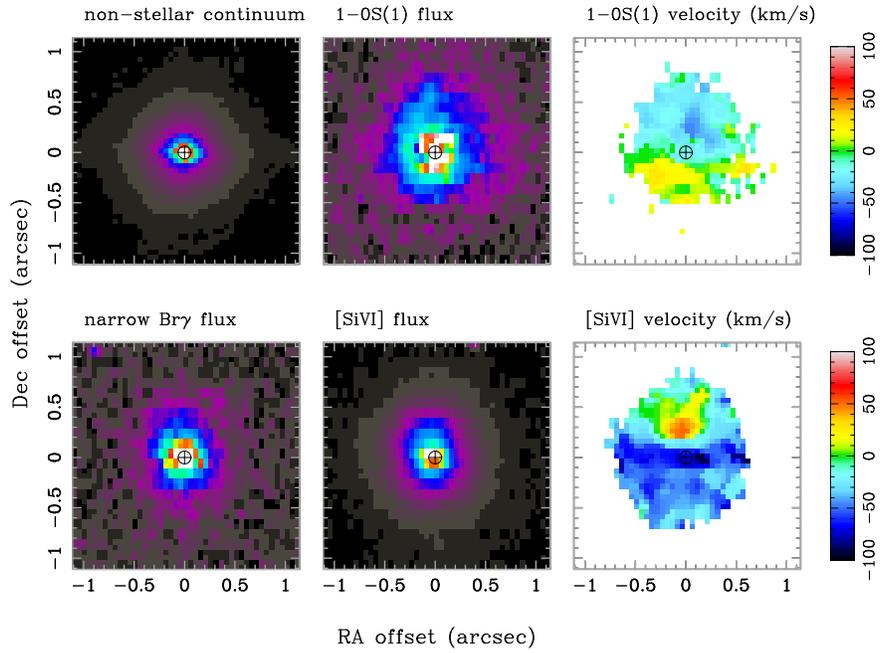} 
\caption{Images of the central 2\arcsec\ of NGC\,3783 (1\arcsec\ = 200\,pc).
Top row from left: non-stellar continuum, H$_2$ 1-0\,S(1) line flux,
H$_2$ 1-0\,S(1) velocity.
Bottom row from left: narrow Br$\gamma$ line flux, [Si{\sc vi}] line
flux, [Si{\sc vi}] velocity.
The Br$\gamma$ velocity field is similar to that of [Si{\sc vi}] and
shows an outflow of $>$50\,km\,s$^{-1}$ to the north.
This is in contrast to the 1-0\,S(1) velocity field which traces
rotation.
}
\label{fig:n3783_maps}
\epsscale{1.0}
\end{figure}


\end{document}